\documentclass[11pt,a4paper]{article}
\pdfoutput=1
\usepackage{cancel}
\usepackage{enumitem}
\usepackage{etex}
\usepackage{amsthm}
\usepackage{geometry}
\usepackage{dcolumn}
\usepackage{epsf}
\usepackage{mathrsfs}
\usepackage{multirow}
\usepackage{booktabs}
\usepackage{tabularx}
\usepackage{array}
\usepackage{slashed}
\usepackage{float}
\usepackage{leftidx}
\usepackage{setspace}
\usepackage{verbatim}
\usepackage{adjustbox}
\usepackage{tikz}
\usepackage[caption=false]{subfig}
\usepackage{arydshln}
\usepackage{jheppub}
\usepackage{shuffle}
\usepackage{amsmath}
\usepackage{cases}

\numberwithin{equation}{section}
\usetikzlibrary{arrows,decorations.markings,shapes.arrows,patterns,positioning}

\newcommand{\bea}{\begin{eqnarray}}
\newcommand{\eea}{\end{eqnarray}}
\newcommand{\bean}{\begin{eqnarray*}}
\newcommand{\eean}{\end{eqnarray*}}
\newcommand{\nn}{\nonumber\\}
\newcommand{\Sl}{\sum\limits}

\def\W #1{\widetilde{#1}}

\def\Label#1{\label{#1}%
  \smash{\hbox to0pt{\raise1ex\hbox{\tiny[#1]}\hss}}}
\def\Label#1{\label{#1}}
\renewcommand{\eqref}[1]{eq.~(\ref{#1})}
\newcommand{\figref}[1]{Fig.~\ref{#1}}

\newcommand{\secref}[1]{section~\ref{#1}}
\newcommand{\appref}[1]{appendix~\ref{#1}}

\def\braket#1{\left\langle #1 \right\rangle}

\def\d{\partial}

\def\vev{\braket}

\def\bvev#1{\left[ #1 \right]}
\def\Spaa{\vev}
\def\Spbb{\bvev}

\def\Sl{\sum\limits}

\newcommand{\ctobedelete}[1]{}
\allowdisplaybreaks

\title{Evaluating EYM amplitudes in four dimensions  by refined graphic expansion}

\author[a]  {Hongxiang Tian \footnote{The unusual ordering of authors is to let authors get proper recognition of contributions under
the practice in China.}} \author[a,b] {Enze Gong} \author[a] {Chongsi Xie} \author[a,c] {Yi-Jian Du\footnote{Corresponding author}}

\affiliation[a]{School of Physics and Technology,
Wuhan University, No.299 Bayi Road, Wuhan 430072, China}

\affiliation[b]{Department of Physics and Astronomy, Uppsala University, 75108 Uppsala, Sweden}

\affiliation[c]{Suzhou Institute of Wuhan University, No.377 Linquan Street, Suzhou, 215123, China}

\emailAdd{thx\underline{~}world@whu.edu.cn}  \emailAdd{Enze.Gong.2675@student.uu.se}

\emailAdd{chongsi.xie@whu.edu.cn} \emailAdd{yijian.du@whu.edu.cn}

\date{\today}
\abstract{The recursive expansion of tree level multitrace Einstein-Yang-Mills (EYM) amplitudes induces a refined graphic expansion, by which any tree-level EYM amplitude can be expressed as a summation over all possible refined graphs. Each graph contributes a unique coefficient as well as a proper combination of color-ordered Yang-Mills (YM) amplitudes. This expansion allows one to evaluate EYM amplitudes through YM amplitudes, the latter have much simpler structures in four dimensions than the former. In this paper, we classify the refined graphs for the expansion of EYM amplitudes into $\text{N}^{\,k}$MHV sectors. Amplitudes in four dimensions, which involve $k+2$ negative-helicity particles,  at most get non-vanishing contribution from graphs in $\text{N}^{\,k'(k'\leq k)}$MHV sectors. By the help of this classification, we evaluate the non-vanishing amplitudes with two negative-helicity particles in four dimensions. We establish a correspondence between the refined graphs for single-trace amplitudes with $(g^-_i,g^-_j)$ or $(h^-_i,g^-_j)$ configuration and the spanning forests of the known Hodges determinant form.  Inspired by this correspondence, we further propose a symmetric formula of double-trace amplitudes with $(g^-_i,g^-_j)$ configuration. By analyzing the cancellation between refined graphs in four dimensions, we prove that any other tree amplitude with two negative-helicity particles has to vanish.
}

\begin{document}
\maketitle  \flushbottom

\section{Introduction}

The expansion of Einstein-Yang-Mills (EYM) \footnote{In this paper, EYM theory always means the theory which also involves antisymmetric B field and dilation, while the external particles of amplitudes can only be gravitons and gluons. } amplitudes in terms of pure Yang-Mills (YM) amplitudes was firstly suggested in \cite{Stieberger:2016lng}, where a tree-level single-trace EYM amplitude with one graviton and $n$ gluons was written as a combination of $n+1$ gluon color-ordered YM amplitudes. This observation was then extended to amplitudes with a few gravitons and/or gluon traces in many literatures, see e.g.,  \cite{Nandan:2016pya,delaCruz:2016gnm,Schlotterer:2016cxa}. A general study of the expansion of single-trace amplitudes can be found in \cite{Fu:2017uzt,Chiodaroli:2017ngp,Teng:2017tbo}. As pointed in \cite{Fu:2017uzt}, a single-trace EYM amplitude could be expanded recursively in terms of EYM amplitudes with fewer gravitons. Thus, the pure-YM expansion can be obtained by applying the recursive expansion iteratively. Along this line, the general formulas for recursive expansion of an arbitrary tree-level multitrace EYM amplitude were established in \cite{Du:2017gnh}.
Further applications and generalizations of this recursive expansion have been investigated, including the symmetry-induced identities \cite{Fu:2017uzt,Du:2017gnh,Hou:2018bwm,Du:2019vzf}, the proof \cite{Du:2018khm} of the equivalence between distinct approaches to amplitudes in nonlinear sigma model \cite{Du:2016tbc,Carrasco:2016ldy,Du:2017kpo}, the construction of polynomial Bern-Carrasco-Johansson (BCJ) \cite{Bern:2008qj} numerators \cite{Fu:2017uzt,Du:2017kpo}, the expansion into BCJ basis \cite{Feng:2020jck,Feng:2019tvb} as well as the generalization to amplitudes in other theories \cite{Plefka:2018zwm,Zhou:2020mvz}.


Among these progresses, a refined graphic rule which conveniently expands an EYM amplitude by summing over a set of so-called \emph{refined graphs} has been invented \cite{Hou:2018bwm,Du:2019vzf}. Each tree graph in this expansion provides a  coefficient (expressed by a product of the Lorentz contractions of external polarizations and/or momenta) as well as a proper combination of color-ordered YM amplitudes. This refined graphic expansion was already shown to be a powerful tool in the study of the relationship between symmetry-induced identities and BCJ relations  \cite{Hou:2018bwm,Du:2019vzf}.

 On the other hand, EYM amplitudes with particular helicity configurations in four dimensions was also studied in many work. By substituting solutions to scattering equations  directly into the Cachazo-He-Yuan formula \cite{Cachazo:2013gna,Cachazo:2013hca,Cachazo:2013iea,Cachazo:2014nsa,Cachazo:2014xea}, the single-trace EYM amplitudes with the maximally-helicity-violating (MHV) configuration, which involves only two negative-helicity particles  $(i^-,j^-)$ ($i,j$ can be either (i) two gluons or (ii) one graviton and one gluon), have been shown to satisfy a compact formula \cite{Du:2016wkt} that is expressed by the well known Hodges determinant \cite{Hodges:2012ym}. Through a spanning forest expansion \cite{Feng:2012sy}, this compact formula was shown (see \cite{Du:2016wkt}) to be equivalent to a generating functional formula which was proposed in earlier work \cite{Selivanov:1997ts,Selivanov:1997aq,Bern:1999bx}. At double-trace level, a symmetric formula for the MHV amplitudes with only external gluons was founded \cite{Cachazo:2014nsa}. In addition, explicit results of five- and six-point examples of double-trace MHV amplitudes with one and two gravitons were also provided in \cite{Cachazo:2014xea}. However, it is still lack of a general symmetric formula of double-trace MHV amplitudes with an arbitrary number of gravitons.

Since  YM amplitudes have much simpler forms than EYM ones, the refined graphic expansion may provide a new approach to the EYM amplitudes in four dimensions. In this paper, we evaluate the EYM amplitudes in four dimensions by the refined graphic rule. To achieve this, we classify refined graphs into $\text{N}^{\,k}$MHV sectors and demonstrate that the $\text{N}^{\,k}$MHV amplitudes with $(k+2)$ negative-helicity particles at most get nonvanishing contributions from graphs in the $\text{N}^{\,k'(k'\leq k)}$MHV sectors. When $k=0$, the nonvanishing MHV amplitudes can only get contribution from the refined graphs in the MHV sector and the corresponding YM amplitudes in the expansion satisfy Parke-Taylor formula \cite{Parke:1986gb}. For single-trace MHV amplitudes with two negative-helicity gluons or those with one-negative helicity graviton and one negative-helicity gluon, a correspondence between the refined graphs and the spanning forests of the Hodges determinant is established. Hence this approach precisely reproduces the known results \cite{Du:2016wkt}. Inspired by this correspondence, we further propose a spanning forest formula of double-trace MHV amplitudes with two negative-helicity gluons, in which the following symmetries are manifest: \emph{(i).} invariance under permutations of gravitons, \emph{(ii).} invariance under exchanging the two traces, \emph{(iii).} the cyclic symmetry of each trace. By an analysis on the cancellation between graphs, we show that all other tree amplitudes with two negative-helicity particles have to vanish.

The structure of this paper is organized as follows. In \secref{sec:MHVsector}, we provide the construction rule for the MHV sector of tree level single- and double-trace EYM amplitudes. By the help of the spinor-helicity formalism \cite{Xu:1986xb} in four dimensions and helpful identities between Parke-Taylor factors, we establish a correspondence between the refined graphs in the MHV sector for single-trace amplitudes and the spanning forests of Hodges determinant in \secref{sec:SingleTrace}. In \secref{sec:DoubleTraceConf1}, a symmetric formula of double-trace MHV amplitudes with two negative-helicity gluons is derived.
Other tree amplitudes with two negative helicity particles are proven to vanish in  \secref{sec:VanishingAmplitude}. This work is summarized in \secref{sec:Conclusions}. The refined graphic rule for the $\text{N}^{\,k}$MHV sector of an $m$-trace amplitude is provided in \appref{app:RefinedNkMHV}. Spinor-helicity formalism and helpful identities are reviewed in \appref{app:SpinorIdentity}.

\subsubsection*{Convention of notations}
Convention of notations is displayed as follows\footnote{Most notations follow from the paper \cite{Du:2019vzf}.}.

{\bf Permutations}: We use boldface Greek letters  $\pmb{\beta}$, $\pmb{\gamma}$, $\pmb{\sigma}$,  or boldface uppercase Latin letters $\pmb{X}$ and $\pmb{Y}$ to denote permutations (or in other words ordered sets). The $i$-th element in a permutation $\pmb{\beta}$ is usually denoted by $\beta_i$, while the position of an element $a\in\pmb{\beta}$ is expressed by $\beta^{-1}(a)$. The condition $\beta^{-1}(a)<\beta^{-1}(b)$ is written as $a\prec b$ in some places for short. The inverse permutation of $\pmb{\beta}$  is denoted by $\pmb{\beta}^{T}$. Shuffle permutations of two ordered sets $\pmb{X}$ and $\pmb{Y}$ are written as $\pmb{X}\shuffle\pmb{Y}$. Number of elements in $\pmb{X}$ is denoted as  $|\pmb{X}|$.

{\bf Gravitons, gluon traces and amplitudes:} We denote gluon traces by boldface numbers $\pmb{1}$, $\pmb{2}$, ...,  or boldface lowercase Latin letters $\pmb{t}$, $\pmb{i}$ ... If a trace $\pmb{i}$ can be written as $\pmb{i}=\{a_i,\pmb{X}_i,b_i,\pmb{Y}_i\}$, we define $\mathsf{KK}[\pmb{i},a_i,b_i]\equiv\pmb{X}_i\shuffle \pmb{Y}_i^T$ and $(-1)^{|\pmb{i},a_i,b_i|}\equiv(-1)^{|\pmb{Y}_i|}$.
A multitrace EYM amplitude is generally denoted by $A(1,2,\ldots, r|\pmb{2}|\ldots|\pmb{m}\Vert\mathsf{H})$, where $\mathsf{H}$ stands for the graviton set with gravitons $h_1$, $h_2$, ..., while $\pmb{1},\pmb{2},\dots,\pmb{m}$ are the gluon traces. Gluons in the trace $\pmb{1}$ are explicitly denoted by $1,\dots,r$. When we study double trace amplitudes, we use $x_1,\dots,x_r$ and $y_1,\dots,y_s$ to denote the gluons in traces $\pmb{1}$ and $\pmb{2}$ respectively. The set of all  gravitons and gluon traces is denoted by ${\pmb{\mathcal{H}}}$.

{\textbf{Refined graphs and spanning forests:}} In this paper, graphs constructed by refined graphic rule are mentioned as \emph{refined graphs} and denoted by  $\mathcal{F}$, while graphs that describe helicity amplitudes in the final formula are mentioned as \emph{spanning forests} and denoted by $\mathcal{G}$. When we fix $1$ and $r$ as the first and the last elements, the set of permutations (of other nodes) established by a refined graph $\mathcal{F}$ is denoted by ${\mathcal{F}}|_1\setminus\{1,r\}$. Components of a graph $\mathcal{F}$ are denoted by $\mathscr{C}_i$, while a chain of components is denoted by $\mathbb{CH}$. We use $\mathsf{R}$ to stands for the reference order of gravitons and traces in the refined graphic rule and use $\mathsf{R}_{\mathscr{C}}$ to denote the reference order of components. The set of refined graphs which involve $m$ traces and belong to the $\text{N}^{\,k}$MHV sector is denoted by $\mathcal{F}\in{\mathcal{F}{[k',m]}}$. Disjoint union of graphs are denoted by $\oplus$.

\section{Sectors of tree-level EYM amplitudes}\label{sec:MHVsector}
%
 \begin{figure}
\centering
\includegraphics[width=0.5\textwidth]{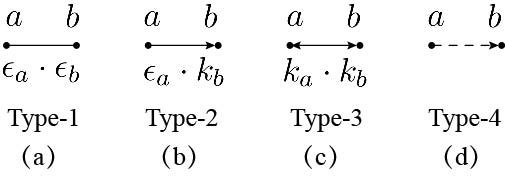}
\caption{Four types of lines in the refined graphic rule for multitrace EYM amplitudes}\label{Fig:LineStyles}
\end{figure}
As pointed out in \cite{Du:2017gnh,Du:2019vzf}, an arbitrary tree level EYM amplitude $A(1,2,\ldots,r|\pmb{2}|\ldots|\pmb{m}\Vert\mathsf{H})$ with the graviton set $\mathsf{H}$ and the gluon traces $\pmb{1}=\{1,2,\ldots, r\},\pmb{2},\dots,\pmb{m}$ can be generally expressed as a combination of  tree level color-ordered pure YM amplitudes
\bea
A\left(1,2,\ldots, r|\pmb{2}|\ldots|\pmb{m}\Vert\mathsf{H}\right)=\Sl_{\mathcal{F}} (-)^{\mathcal{F}}\mathcal{C}^{\mathcal{F}}\biggl[\,\Sl_{\pmb{\sigma}\in{\mathcal{F}|_1\setminus\{1,r\}}}A\big(1,\pmb{\sigma},r\big)\,\biggr],\Label{Eq:PureYMExpansion}
\eea
where  we have summed over all possible connected tree graphs $\mathcal{F}$ which are constructed according to the refined graphic rule \cite{Du:2017gnh,Du:2019vzf}.
A tree graph $\mathcal{F}$ defines a coefficient $\mathcal{C}^{\mathcal{F}}$ as well as a combination of color-ordered YM amplitudes $A\big(1,\pmb{\sigma},r\big)$. The polarization tensor $\epsilon_h^{\mu\nu}$ of a graviton splits into two polarization vectors $\epsilon_h^{\mu}\epsilon_h^{\nu}$. One of them is absorbed into the coefficient $\mathcal{C}^{\mathcal{F}}$, the other is considered as the polarization vector of a gluon which is involved in the pure YM amplitude $A\big(1,\pmb{\sigma},r\big)$. In each graph $\mathcal{F}$, all gluons and gravitons are expressed by nodes.  As shown by \figref{Fig:LineStyles}, distinct types of lines between nodes are introduced: (a). A type-1 line represents an $\epsilon\cdot\epsilon$ factor between two gravitons; (b). A type-2 line denotes an $\epsilon\cdot k$ factor ($k^{\mu}$ denotes the momentum of a node) between a graviton and a graviton/gluon; (c). A type-3 line stands for a $k\cdot k$ factor between any two nodes; (d). A type-4 line is  introduced to record the relative order of two adjacent gluons in a gluon trace but does not contribute any nontrivial kinematic factor. The expansion coefficient $\mathcal{C}^{\mathcal{F}}$ is then given by the product of factors defined by all lines in $\mathcal{F}$, while a proper sign $(-)^{\mathcal{F}}$ is also defined (see \appref{app:RefinedNkMHV}).
Permutations $\pmb{\sigma}\in{\mathcal{F}|_1\setminus\{1,r\}}$ established by the graph $\mathcal{F}$ are given by the following two steps: (i). If $a$ and $b$ are two adjacent nodes which satisfy the condition that $a$ is nearer to the gluon $1$ than $b$,  we have $\sigma^{-1}(a)<\sigma^{-1}(b)$ where $\sigma^{-1}(a)$ and $\sigma^{-1}(b)$ respectively denote the positions of $a$ and $b$ in $\pmb{\sigma}$.\footnote{Supposing that the position of $a$ in $\pmb{\sigma}$ is $j$, we have $a=\sigma_j\equiv{\sigma}(j)$, hence it is reasonable to define $j={\sigma}^{-1}(a)$.} (ii). If two subtree structures are attached to a same node, we shuffle the corresponding permutations together such that the permutation in each subtree is preserved.

In order to analyze  helicity amplitudes with $m$ traces in four dimensions, we denote the set of all graphs involving $k$ type-1 lines (i.e.  $\epsilon\cdot\epsilon$ lines) and $m$ gluon traces by $\mathcal{F}{[k,m]}$. We further define the $\text{N}^{\,k}$MHV sector $I[k,m]$ ($k\geq 0$) of the expansion (\ref{Eq:PureYMExpansion}) by the total contribution of graphs in $\mathcal{F}{[k,m]}$ as follows:
\bea
I\left[k,m\right]\equiv\Sl_{\mathcal{F}\in\mathcal{F}{\left[k,m\right]}}(-)^{\mathcal{F}}\mathcal{C}^{\mathcal{F}}\,\Sl_{\pmb{\sigma}\in{\mathcal{F}|_1\setminus\{1,r\}}}A\bigl(1,\pmb{\sigma},r\bigr).
\eea
The expansion (\ref{Eq:PureYMExpansion}) is then given by summing over all possible $k$
 \bea
A\left(1,2,\ldots, r|\pmb{2}|\ldots|\pmb{m}\Vert\mathsf{H}\right)=\Sl_{k}\,I\left[k,m\right].\Label{Eq:PureYMExpansion1}
\eea
The general refined graphic rule for constructing the $\text{N}^{\,k}$MHV sector is provided in \appref{app:RefinedNkMHV}. \emph{In the following, we illustrate that an $\text{N}^{\,k}$MHV amplitude in four dimensions can at most get nonzero contribution from graphs in the $\text{N}^{\,k'(k'\leq k)}$MHV sector.}

In four dimensions, both the momentum of a graviton/gluon and the (half) polarization of a graviton can be expressed according to the spinor-helicity formalism \cite{Xu:1986xb} (see \appref{app:SpinorIdentity}). The Lorentz contractions between momenta and/or polarizations are then expressed by  spinor products, see \eqref{Eq:LorentzContractions}. In this paper, the reference momenta of all those gravitons with the same helicity are chosen as the same one. With this choice of gauge, the contractions between the (half) polarizations of any two gravitons with the same helicity must vanish. The only possible remaining contractions between `half' polarizations must have the form $\epsilon_{h_i}^{+}\cdot\epsilon_{h_j}^{-}$. For an $m$-trace $\text{N}^{\,k}$MHV amplitude with $l$ negative-helicity gravitons and $k-l+2$ negative-helicity gluons, the number of factors $\epsilon_{h_i}^{+}\cdot\epsilon_{h_j}^{-}$ in the coefficient $\mathcal{C}^{\mathcal{F}}$ is at most $l$. This implies only those graphs $\mathcal{F}\in{\mathcal{F}{[k',m]}}$, which contain $k'$ ($k'\leq l$) type-1 lines, may provide nonvanishing contributions to an $\text{N}^{\,k}$MHV amplitude in four dimensions. The number of type-1 lines
can be further reduced when the following two facts are considered:
\begin{itemize}
\item [(i).] It has been known that an EYM amplitude where all the negative-helicity particles are gravitons (in other words all gluons carry the same helicity) must vanish. Thus amplitudes in the case of $l=k+2$, for which a graph at most involves  $l=k+2$ nontrivial type-1 lines, have to vanish.
\item [(ii).] When the reference momentum of all positive-helicity gravitons are chosen as the momentum of one of the negative-helicity gravitons, say $h_i$, the maximal number of type-1 lines in a graph is further reduced by one because $\epsilon^{-}_{h_i}(p)\cdot \epsilon^{+}_{h_j}(k_{h_i})=0$. Thus amplitudes with $l=k+1$ can only get nonzero contributions from graphs $\mathcal{F}\in{\mathcal{F}{[k',m]}}$ ($k'\leq k$).    \end{itemize}
Therefore, the nonvanishing EYM amplitudes with $k+2$ negative-helicity particles can at most get nonzero contributions from the graphs in the $\text{N}^{\,k'(k'\leq k)}$MHV sector.

In the current paper, we study the single- and double-trace amplitudes with only two negative helicity particles. The helicity configurations can be classified into three categories $(g_i^-,g_j^-)$, $(h_i^-,g_j^-)$ and $(h_i^-,h_j^-)$ which correspond to EYM amplitudes with (i). two negative-helicity gluons, (ii). one negative-helicity gluon and one negative-helicity graviton and (iii). two negative-helicity gravitons. As stated before, the last configuration has to vanish, while the first two configurations only get contributions from  graphs in the MHV sector. In the coming sections, we show that the MHV sectors of single-trace amplitudes with the $(g_i^-,g_j^-)$ and $(h_i^-,g_j^-)$ configurations precisely match with the corresponding spanning forests of the Hodges determinant form in four dimensions. By generalizing this discussion  to double-trace amplitudes, we establish a spanning forest formula for double-trace amplitudes with the  $(g_i^-,g_j^-)$ configuration. Single-trace amplitudes with the $(h_i^-,h_j^-)$ configuration, double-trace amplitudes with $(h_i^-,g_j^-)$ and $(h_i^-,h_j^-)$ configurations as well as  all $m(m>3)$-trace amplitudes with two negative-helicity particles  will be proven to vanish.

\begin{figure}
\centering
\includegraphics[width=0.74\textwidth]{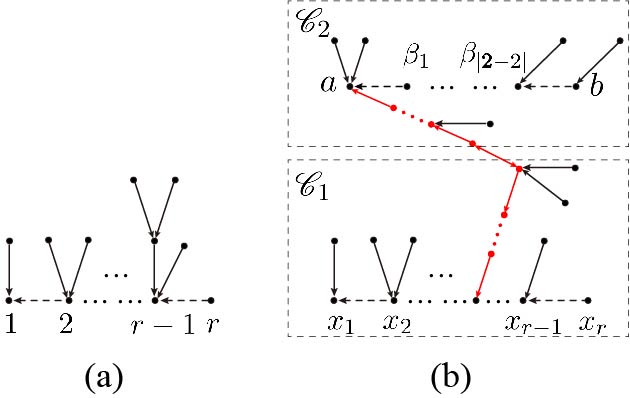}
\caption{(a): A typical graph for the MHV sector of single-trace EYM amplitudes.  (b): A typical graph for the MHV sector of double-trace EYM amplitudes.}\label{Fig:MHVSingleDouble}
\end{figure}

Before proceeding, we explicitly display the expansions of the single- and double-trace MHV amplitudes by graphs in the corresponding MHV sectors. The general construction rule of the $\text{N}^{\,k}$MHV sector can be found in \appref{app:RefinedNkMHV}.

\subsection{The expansion of single-trace MHV amplitudes}\label{sec:ExpansionSingleTrace}

Following the above discussion,  a single-trace MHV amplitude $A^{(i,j)}(1,2,\ldots, r\Vert\mathsf{H})$ (where the two negative helicity particles $(i,j)$ can be either $(g_i,g_j)$ or $(h_i,g_j)$) is given by
 \bea
 A^{(i,j)}(1,2,\ldots, r\Vert\mathsf{H})=\Sl_{\mathcal{F}\in\mathcal{F}{[0,1]}}(-)^{\mathcal{F}}\mathcal{C}^{\mathcal{F}}A^{(i,j)}(1,\pmb{\sigma},r)=\Sl_{\mathcal{F}\in\mathcal{F}{[0,1]}} \mathcal{C}^{\mathcal{F}}\Biggl[\,\Sl_{\pmb{\sigma}\in{\mathcal{F}\,|_1\setminus\{1,r\}}}\frac{\Spaa{i,j}^4}{(1,\pmb{\sigma},r)}\Biggr],\Label{Eq:SingleMHVYMExpansion1}
\eea
in which all graphs $\mathcal{F}$ in the MHV sector for the single-trace amplitude have been summed over, while each pure-YM amplitude $A^{(i,j)}(1,\pmb{\sigma},r)$ involving two negative-helicity particles $(i,j)$  is further expressed by the Parke-Taylor formula.  Here, $(1,2,\dots,r)$ denotes $\Spaa{1,2}\Spaa{2,3}\dots\Spaa{r,1}$ for short.

Graphs $\mathcal{F}\in\mathcal{F}{[0,1]}$ in \eqref{Eq:SingleMHVYMExpansion1} are obtained by the general construction rule in \appref{app:RefinedNkMHV}. To be specific, the general pattern of a graph $\mathcal{F}\in\mathcal{F}[0,1]$ is given by connecting gravitons to gluons in $\pmb{1}\setminus\{r\}$ via type-2 lines (i.e., $\epsilon\cdot k$ lines) whose arrows point towards the trace $\pmb{1}$, as shown by \figref{Fig:MHVSingleDouble} (a). According to  \appref{app:RefinedNkMHV}, the sign $(-)^{\mathcal{F}}$ associating to any graph $\mathcal{F}$ of this pattern is  $+1$.

\subsection{The expansion of double-trace MHV amplitudes}\label{sec:ExpansionDoubleTrace}


Suppose the two gluon traces are $\pmb{1}=\{x_1,\dots,x_r\}$, $\pmb{2}=\{y_1,\dots,y_s\}$\footnote{Here, we slightly change the labels of gluons so that the two traces are on an equal footing in the final result.}, the graviton set is $\mathsf{H}$ and the two negative-helicity particles are $(i,j)$ ($(i,j)$  can be either $(g_i,g_j)$ or $(h_i,g_j)$). A double-trace amplitude $A^{(i,j)}(x_1,\ldots, x_r|y_1,\ldots,y_s\Vert\mathsf{H})$ is expanded as
\bea
A^{(i,j)}(x_1,\ldots, x_r|y_1,\ldots,y_s\Vert\mathsf{H})=\Sl_{\mathcal{F}\in\mathcal{F}{[0,2]}} (-)^{\mathcal{F}}\mathcal{C}^{\mathcal{F}}\Biggl[\,\Sl_{\pmb{\sigma}\in{\mathcal{F}\,|_{x_1}\setminus\{x_1,x_r\}}}\frac{\Spaa{i,j}^4}{(x_1,\pmb{\sigma},x_r)}\Biggr],\Label{Eq:DoubleMHVYMExpansion1}
\eea
where we have summed over all possible graphs $\mathcal{F}$ in the MHV sector of double-trace amplitudes.

As stated in \appref{app:RefinedNkMHV}, graphs $\mathcal{F}\in\mathcal{F}{[0,2]}$ have the general pattern \figref{Fig:MHVSingleDouble} (b) where two components $\mathscr{C}_1$ and $\mathscr{C}_2$, which respectively contains the two traces $\pmb{1}$ and $\pmb{2}$, are connected by a type-3 line (i.e. the $k\cdot k$ line). Such a graph has the following pattern:
\begin{itemize}
 \item Gluons of the trace $\pmb{1}$ is arranged in the normal order $x_1,\dots,x_r$. Assuming the trace $\pmb{2}$ can be written as $\pmb{2}=\{a,\pmb{X},b,\pmb{Y}\}$ where $a,b\in\pmb{2}$, all gluons in  $\pmb{2}$ are arranged in an order $a,{\beta}_1,\dots,\beta_{|\pmb{2}|-2},b$ where $\{{\beta}_1,\dots,\beta_{|\pmb{2}|-2}\}\in\pmb{X}\shuffle \pmb{Y}^T$.

 \item Each of $\mathscr{C}_1$ and $\mathscr{C}_2$ in \figref{Fig:MHVSingleDouble} (b) is constructed by connecting gravitons therein to gluons in the corresponding trace  (except the last element $x_r$ in the trace $\pmb{1}$) via type-2 lines. The arrow of any type-2 line should point towards the direction of the corresponding trace.

 \item The two end nodes $x$ and $y$ of the type-3 line $k_x\cdot k_y$ respectively belong to $\mathscr{C}_1$ and $\mathscr{C}_2$. The node $x$ satisfies $x\in \mathscr{C}_1\setminus\{x_r\}$
 and the node $y$ is either the node $a$ or a node which is connected to $a$ via only type-3 lines.

 \item  The sign $(-)^{\mathcal{F}}$ associating to a graph $\mathcal{F}$ is given by $(-)^{\mathcal{F}}\equiv(-1)^{\mathcal{N}\left(\mathcal{F}\right)+|\pmb{2},a,b|}$,
where $\mathcal{N}\left(\mathcal{F}\right)$ is the total number of arrows pointing away from $x_1$, $|\pmb{2},a,b|$ denotes the number of elements in the $\pmb{Y}$ set if the trace $\pmb{2}$ can be written as $\pmb{2}=\{a,\pmb{X},b,\pmb{Y}\}$ for a given $a\neq b$ ($a\in\pmb{2}$).

\end{itemize}
The summation over all graphs $\mathcal{F}\in\mathcal{F}{[0,2]}$ in \eqref{Eq:DoubleMHVYMExpansion1}  then means that we should sum over (i). all possible graphs constructed by the above method for a given $a\neq b$ ($b\in\pmb{2}$ is always fixed) and a given $\pmb{\beta}=\{{\beta}_1,\dots,\beta_{|\pmb{2}|-2}\}$, (ii). all possible $\pmb{\beta}$ for a given $a$, (iii). all possible $a\in\pmb{2}$, ($a\neq b$).

\begin{figure}
\centering
\includegraphics[width=0.28\textwidth]{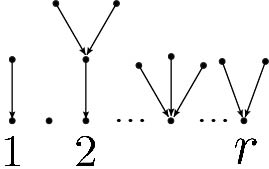}
\caption{A typical spanning forest for single-trace MHV amplitude. All positive-helicity gravitons and gluons are treated as nodes. All trees in the forest are rooted at gluons. }\label{Fig:SpanningForestSingle}
\end{figure}
%

\section{Single-trace MHV amplitudes with $(g_i^-,g_j^-)$ and $(h_i^-,g_j^-)$}\label{sec:SingleTrace}
%

In this section, we evaluate the single-trace MHV amplitudes $A^{(i,j)}(1,2,\ldots, r\Vert\mathsf{H})$ with the helicity configurations $(g^-_i,g^-_j)$ and $(h^-_i,g_j^-)$ by the expansion (\ref{Eq:SingleMHVYMExpansion1}). We prove that \eqref{Eq:SingleMHVYMExpansion1} precisely gives the spanning forest form
\bea
\boxed{A^{(i,j)}(1,2,\ldots, r\Vert\mathsf{H})=(-1)^{N_{h^-}}\frac{\Spaa{i,j}^4}{\big(1,2,\dots,r\bigr)}\left[\,\Sl_{\mathcal{G}}\,\prod\limits_{e(x,y)\in \mathcal{E}(\mathcal{G})}\frac{\Spaa{y,\xi}\Spaa{y,\eta}\Spbb{y,x}}{\Spaa{x,\xi}\Spaa{x,\eta}\Spaa{y,x}}\,\right]}, \Label{Eq:SingleTraceSpanningTree}
\eea
which is the spanning forest expansion \cite{Feng:2012sy} of the known formula (\ref{Eq:MHVYM}) (see \cite{Du:2016wkt}) that is expressed by Hodges determinant \cite{Hodges:2012ym}. In the above expression, we summed over all possible spanning forests $\mathcal{G}$ (see \figref{Fig:SpanningForestSingle}) where all elements in $\{1,\dots,r\}\cup \mathsf{H}^+$ ($\mathsf{H}^+$ is the set of positive-helicity gravitons)  were treated as nodes, while the gluons $1,\dots,r$ were considered as roots. The prefactor $(-1)^{N_{h^-}}$ is $+1$ for the  $(g^-_i,g^-_j)$ configuration and $-1$ for the  $(h^-_i,g_j^-)$ configuration. For a given forest $\mathcal{G}$, $\mathcal{E}(\mathcal{G})$ denotes the set of edges and $e(x,y)$ denotes an edge between nodes $x$ and $y$. For an edge $e(x,y)$,  $y$ is always supposed to be nearer to root than $x$ and the edge is dressed by an arrow pointing towards $y$. Each $e(x,y)$ in \eqref{Eq:SingleTraceSpanningTree} is accompanied by a factor
\bea
\frac{\Spaa{y,\xi}\Spaa{y,\eta}\Spbb{y,x}}{\Spaa{x,\xi}\Spaa{x,\eta}\Spaa{y,x}},
\eea
where $\xi$ and $\eta$ are two arbitrarily chosen reference spinors which cannot make the expression divergent.  In the following, we prove \eqref{Eq:SingleTraceSpanningTree} with $(g^-_i,g^-_j)$ and  $(h^-_i,g_j^-)$ configurations in turn.
%

\subsection{Single-trace amplitudes with $(g^-_i,g^-_j)$ configuration}

For a single-trace MHV amplitude with the $(g^-_i,g^-_j)$ configuration, all gravitons carry positive helicity. As shown in \secref{sec:MHVsector}, the reference momenta of all gravitons are taken as the same one, say $\xi^{\mu}$. According to spinor helicity formalism, the Lorentz contraction $\epsilon^+_{h}\cdot k_l$ between the half polarization $\epsilon^{+\mu}_{h}$ of a graviton $h$ and an arbitrary momentum $k^{\mu}_l$ becomes
\bea
\epsilon^+_{h_a}(\xi)\cdot k_l=\frac{\Spaa{\xi,l}\Spbb{l,h_a}}{\Spaa{\xi,h_a}}.\Label{Eq:type-2Line}
\eea
Therefore, the coefficient $\mathcal{C}^{\mathcal{F}}$ in \eqref{Eq:SingleMHVYMExpansion1} can be conveniently expressed as
\bea
\mathcal{C}^{\mathcal{F}}=\mathcal{C}^{\mathcal{F}\setminus\{h_a\}}\,\frac{\Spaa{\xi,l}\Spbb{l,h_a}}{\Spaa{\xi,h_a}} \Label{Eq:LorentzContraction}
\eea
for a leaf (i.e. an outmost graviton) $h_a\in\mathcal{F}$. Here  $\mathcal{F}\setminus\{h_a\}$ denotes the subgraph that is obtained from $\mathcal{F}$, by removing $h_a$ and the edge attached to it. The node $l$ in \eqref{Eq:LorentzContraction} is supposed to be the neighbour of $h_a$.
 \begin{figure}
\centering
\includegraphics[width=0.97\textwidth]{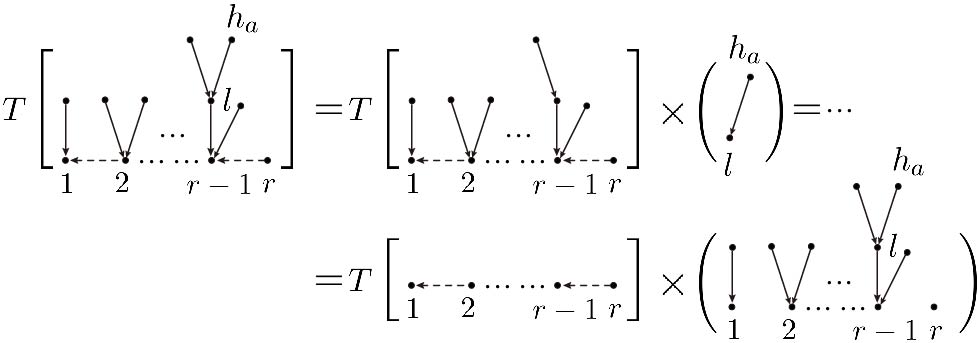}
\caption{Evaluating a single-trace amplitude with the $(g_i^-,g_j^-)$ configuration.}\label{Fig:SingleTraceRecursion}
\end{figure}
On another hand, the summation over all permutations $\pmb{\sigma}\in{\mathcal{F}\,|_1\setminus\{1,r\}}$ which are established by the graph  $\mathcal{F}$ in \eqref{Eq:SingleMHVYMExpansion1} can be achieved by the following steps: (1). summing over all possible permutations $\pmb{\gamma}\in\mathcal{F}\big|_{1}\setminus\{1,r,h_a\}$  established by the graph $\mathcal{F}\setminus\{h_a\}$; (2).  summing over all possible permutations $\pmb{\gamma}\shuffle\{h_a\}|_{l\prec h_a}$ \footnote{Here we write the constraint condition $\sigma^{-1}(l)< \sigma^{-1}(h_a)$ by $l\prec h_a$ for short.} for a given $\pmb{\gamma}$ in the previous step. Thus the  summation in square brackets of \eqref{Eq:SingleMHVYMExpansion1} is rearranged into
\bea
\Sl_{\pmb{\gamma}\in\mathcal{F}|_{1}\setminus\{1,r,h_a\}}\,\Sl_{\pmb{\sigma}\in\pmb{\gamma}\,\shuffle\{h_a\}|_{l\prec h_a}}\,\frac{\Spaa{g_i,g_j}^4}{\big(1,\pmb{\sigma},r\big)},\Label{Eq:SumPTSingle}
\eea
which can be further simplified  by the help of the identity (\ref{Eq:PTRelation1}),  as follows
\bea
\Biggl[\,\Sl_{\pmb{\gamma}\in\mathcal{F}|_{1}\setminus\{1,r,h_a\}}\,\frac{\Spaa{g_i,g_j}^4}{\big(1,\pmb{\gamma},r\big)}\,\Biggr]\,\frac{\Spaa{l,r}}{\Spaa{l,h_a}\Spaa{h_a,r}}.\Label{Eq:SumPTSingle1}
\eea
When substituting \eqref{Eq:SumPTSingle1} and \eqref{Eq:LorentzContraction} into \eqref{Eq:SingleMHVYMExpansion1} for a given graph $\mathcal{F}$, we arrive
\bea
T[\mathcal{F}]&\equiv&\mathcal{C}^{\mathcal{F}}\Biggl[\,\Sl_{\pmb{\sigma}\in{\mathcal{F}\,|_1\setminus\{1,r\}}}\frac{\Spaa{g_i,g_j}^4}{(1,\pmb{\sigma},r)}\,\Biggr]\nn
&=&\Biggl[\,\mathcal{C}^{\mathcal{F}\setminus\{h_a\}}\Sl_{\pmb{\gamma}\in\mathcal{F}|_{1}\setminus\{1,r,h_a\}}\frac{\Spaa{g_i,g_j}^4}{\big(1,\pmb{\gamma},r\big)}\,\Biggr]
\,\frac{\Spaa{l,r}\Spaa{l,\xi}\Spbb{l,h_a}}{\Spaa{h_a,r}\Spaa{h_a,\xi}\Spaa{l,h_a}}\nn
&=&T[\mathcal{F}\setminus\{h_a\}]\,\frac{\Spaa{l,\xi}\Spaa{l,r}\Spbb{l,h_a}}{\Spaa{h_a,\xi}\Spaa{h_a,r}\Spaa{l,h_a}},\Label{Eq:RecursionT}
\eea
where the expression in the square brackets on the second line is just the contribution of the graph $\mathcal{F}\setminus\{h_a\}$.
Eq. (\ref{Eq:RecursionT}) is shown by the first equality of \figref{Fig:SingleTraceRecursion},  while we introduce an arrowed edge $e(h_a,l)$ pointing towards $l$ to denote the factor
\bea
\frac{\Spaa{l,\xi}\Spaa{l,r}\Spbb{l,h_a}}{\Spaa{h_a,\xi}\Spaa{h_a,r}\Spaa{l,h_a}}.\Label{Eq:Edge1}
\eea
Apparently, the removed line between $h_a$ and $l$ in the graph $\mathcal{F}$ is transformed into the edge $e(h_a,l)$ with the new meaning \eqref{Eq:Edge1} in this step.

The equation (\ref{Eq:RecursionT}) establishes a relation between contributions of a graph $\mathcal{F}$ and the subgraph with one gravion extracted out, hence can be applied iteratively. To be specific, we can further pick out a leaf, say $h_b$, from the graph $\mathcal{F}\setminus\{h_a\}$. According to \eqref{Eq:RecursionT}, $T[\mathcal{F}\setminus\{h_a\}]$ can be written as
\bea
T[\mathcal{F}\setminus\{h_a\}]=T\left[\mathcal{F}\setminus\{h_a,h_b\}\right]\frac{\Spaa{l',\xi}\Spaa{l',r}\Spbb{l',h_b}}{\Spaa{h_b,\xi}\Spaa{h_b,r}\Spaa{l',h_b}},
\eea
where  $h_b$ is supposed to be connected to the node $l'$ via a type-2 line.
Repeating the above steps until all gravitons have been extracted out, we get the final expression which consists of the following two factors (see the last term in \figref{Fig:SingleTraceRecursion}):
(i). the Parke-Taylor factor corresponding to the gluon trace $\pmb{1}$, (ii). the coefficient characterized by the forest which has the same structure with that in the original graph $\mathcal{F}$ but with the new meaning of the edges, i.e. (\ref{Eq:Edge1}). When all possible graphs $\mathcal{F}\in \mathcal{F}[0,1]$ are summed over, the full single-trace MHV amplitude \eqref{Eq:SingleMHVYMExpansion1} with the $(g_i^-,g_j^-)$ configuration is then written as
\bea
\frac{\Spaa{g_i,g_j}^4}{\big(1,2,\dots,r\bigr)}\Sl_{\mathcal{G}'}\,\prod\limits_{e(x,y)\in \mathcal{E}(\mathcal{G}')}\frac{\Spaa{y,\xi}\Spaa{y,r}\Spbb{y,x}}{\Spaa{x,\xi}\Spaa{x,r}\Spaa{y,x}}.
\eea
In the above expression, we have summed over all possible spanning forests $\mathcal{G}'$ where leaves and internal nodes are gravitons, gluons $1,2,\dots,r-1$ are roots (recalling that in a graph $\mathcal{F}$ (see \eqref{Eq:SingleMHVYMExpansion1}) the last gluon $r$ cannot be connected by a type-2 line). The summation over $\mathcal{G}'$  can be further extended to a summation over all forests $\mathcal{G}$ rooted at gluons $1,2,\dots,r$ because
\bea
\frac{\Spaa{r,\xi}\Spaa{r,r}\Spbb{r,h_a}}{\Spaa{h_a,\xi}\Spaa{h_a,r}\Spaa{l,h_a}}=0.
\eea
Therefore, we get the expression (\ref{Eq:SingleTraceSpanningTree}) with the choice of  reference spinor $\eta=r$. Since the RHS of \eqref{Eq:SingleTraceSpanningTree} can be  expressed by the Hodges determinant which is independent of the choice of the reference spinors $\xi$ and $\eta$ \cite{Feng:2012sy,Du:2016wkt}), the proof of \eqref{Eq:SingleTraceSpanningTree} has been completed.

 \begin{figure}
\centering
\includegraphics[width=0.6\textwidth]{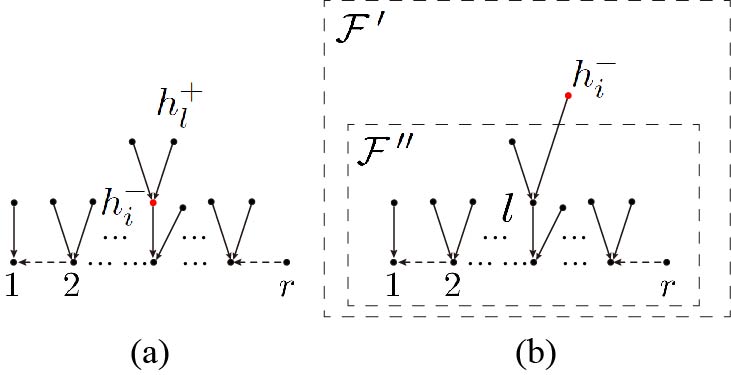}
\caption{Typical graphs for the $(h^-_i,g^-_j)$ configuration. In (a) the negative-helicity graviton $h_i$ is the end node of a type-2 line. In (b), $h_i$ plays as a leaf (i.e. an outmost node).}\label{Fig:SingleTraceConfiguration2}
\end{figure}

\subsection{Single-trace amplitudes with $(h^-_i,g^-_j)$ configuration}\label{sec:SingleTraceConf2}

For the $(h^-_i,g^-_j)$ configuration, there is one negative-helicity graviton $h_i$ and one negative-helicity gluon $g_j$. As stated in \secref{sec:MHVsector},  the reference momenta of all positive-helicity gravitons are  chosen as $k^{\mu}_{h_i}$. Consequently,
\bea
\epsilon_{h_l}^+(k_{h_i})\cdot k_{h_i}=0.
\eea
This indicates those graphs $\mathcal{F}$, where the negative-helicity graviton $h_i$ is the ending node of a type-2 line (as shown by \figref{Fig:SingleTraceConfiguration2} (a)), have to vanish. In other words, $h_i$ in a nonvanishing graph can only be a leaf, as shown by \figref{Fig:SingleTraceConfiguration2} (b). Thus the summation over graphs $\mathcal{F}\in\mathcal{F}{[0,1]}$ in \eqref{Eq:SingleMHVYMExpansion1} becomes a summation $\sum_{\mathcal{F}\,'}$ over all graphs $\mathcal{F}\,'$  with the structure  \figref{Fig:SingleTraceConfiguration2} (b). This summation can be further arranged as $\sum_{\mathcal{F}\,'}\to\sum_{\mathcal{F}\,''}\,\sum_{l\in \mathcal{F}\,''\setminus\{r\}}$,
where $\mathcal{F}\,''$ denote those subgraphs (of graphs $\mathcal{F}'$) which involve only positive-helicity gravitons and all gluons, while $l$ denotes the neighbour of $h_i$ in the original graph $\mathcal{F}'$  (see \figref{Fig:SingleTraceConfiguration2} (b)).
Meanwhile, the coefficient $\mathcal{C}^{\mathcal{F}\,'}$ is factorized as
\bea
\mathcal{C}^{\mathcal{F}\,'}=\mathcal{C}^{\mathcal{F}\,''}\big(\epsilon^-_{h_i}\cdot k_l\big)=\mathcal{C}^{\mathcal{F}\,''}\frac{\Spbb{\eta,l}\Spaa{l,h_i}}{\Spbb{\eta,h_i}},
\eea
where $\eta^{\mu}$ is the reference momentum of the negative-helicity graviton $h_i$. The permutations $\pmb{\sigma}\in \mathcal{F}\,'|_1\setminus\{1,r\}$ can be reexpressed as
\bea
\pmb{\sigma}\in \pmb{\gamma}\shuffle\{h_i\}|_{l\prec h_i}, \text{~where~} \pmb{\gamma}\in \mathcal{F}\,''|_1\setminus\{1,r\}.
\eea
Altogether, the single-trace MHV amplitude $A^{(h_i,g_j)}(1,2,\ldots, r\Vert\mathsf{H})$ in \eqref{Eq:SingleMHVYMExpansion1} becomes
\bea
\Sl_{\mathcal{F}\,''}\mathcal{C}^{\mathcal{F}\,''}\Sl_{l\in \mathcal{F}\,''\setminus\{r\}}\,\frac{\Spbb{\eta,l}\Spaa{l,h_i}}{\Spbb{\eta,h_i}}\Biggl[\,\Sl_{\pmb{\gamma}\in \mathcal{F}\,''|_1\setminus\{1,r\}}\,\Sl_{\pmb{\sigma}\in \pmb{\gamma}\,\shuffle\{h_i\}|_{l\prec h_i}}\,\frac{\Spaa{h_i,g_j}^4}{(1,\pmb{\sigma},r)}\Biggr].\Label{Eq:SingleMHVYMExpansion2}
\eea
When the identity (\ref{Eq:PTRelation1}) is applied, the last summation in the square brackets gives
\bea
\Sl_{\pmb{\sigma}\in \pmb{\gamma}\,\shuffle\{h_i\}|_{l\prec h_i}}\,\frac{\Spaa{h_i,g_j}^4}{(1,\pmb{\sigma},r)}=\frac{\Spaa{h_i,g_j}^4}{\big(1,\pmb{\gamma},r\big)} \frac{\Spaa{l,r}}{\Spaa{l,h_i}\Spaa{h_i,r}},
\eea
Here, the first factor is a Parke-Taylor factor which does not involve the graviton $h_i$ in the denominator, while the second factor, $\frac{\Spaa{l,r}}{\Spaa{l,h_i}\Spaa{h_i,r}}$, depends on the choice of $l$ and is independent of $\pmb{\gamma}$. Thus the summation of the factors depending on $l$ in \eqref{Eq:SingleMHVYMExpansion2}
is given by
\bea
\Sl_{l\in \mathcal{F}\,''\setminus\{r\}}\frac{\Spbb{\eta,l}\Spaa{l,h_i}}{\Spbb{\eta,h_i}} \cdot \frac{\Spaa{l,r}}{\Spaa{l,h_i}\Spaa{h_i,r}}=\Sl_{l\in \mathcal{F}\,''\setminus\{r\}}\frac{\Spaa{l,r}\Spbb{\eta,\,l}}{\Spaa{h_i,r}\Spbb{\eta,h_i}}=-\frac{\Spaa{r,r}\Spbb{\eta,\,r}}{\Spaa{h_i,r}\Spbb{\eta,h_i}}-\frac{\Spaa{h_i,r}\Spbb{\eta,h_i}}{\Spaa{h_i,r}\Spbb{\eta,h_i}}=-1,\label{Eq:hiProperty1}
\eea
where momentum conservation and the antisymmetry of spinor products have been applied. After this simplification, the amplitude (\ref{Eq:SingleMHVYMExpansion2}) turns into
\bea
A^{(h_i,g_j)}(1,2,\ldots, r\Vert\mathsf{H})=(-1)
\Sl_{\mathcal{F}\,''}\mathcal{C}^{\mathcal{F}\,''}\Biggl[\,\Sl_{\pmb{\gamma}\in \mathcal{F}\,''|_1\setminus\{1,r\}}\,\frac{\Spaa{h_i,g_j}^4}{(1,\pmb{\gamma},r)}\Biggr],
\eea
in which, $\mathcal{F}\,''$ are those tree graphs in the MHV sector with the negative-helicity graviton $h_i$ deleted. Up to an overall factor, the above expression exactly has the same pattern with the $(g^-_i,g^-_j)$ case (i.e. \eqref{Eq:SingleMHVYMExpansion1} with the $(g^-_i,g^-_j)$ configuration) in which all gravitons have positive helicity. Thus, following a parallel discussion with the $(g^-_i,g^-_j)$ case, we immediately arrive
\bea
A^{(h_i,g_j)}(1,2,\ldots, r\Vert\mathsf{H})=(-1)\frac{\Spaa{h_i,g_j}^4}{\big(1,2,\dots,r\bigr)}\left[\,\Sl_{\mathcal{G}}\,\prod\limits_{e(x,y)\in \mathcal{E}(\mathcal{G})}\frac{\Spaa{y,h_i}\Spaa{y,r}\Spbb{y,x}}{\Spaa{x,h_i}\Spaa{x,r}\Spaa{y,x}}\,\right], \Label{Eq:SingleTraceSpanningTree3}
\eea
where we have summed over all spanning forests $\mathcal{G}$ with the node set $\{1,\dots,r\}\cup \mathsf{H}^+$ and roots $1$, \dots, $r$. This expression is identical with \eqref{Eq:SingleTraceSpanningTree} when choosing $\xi=h_i$ and $\eta=r$. Since \eqref{Eq:SingleTraceSpanningTree} is independent of the choice of $\xi$ and $\eta$  \cite{Du:2016wkt}, the proof for the $(h^-_i,g^-_j)$ configuration has been completed.

\subsection{Comments}

Now we summarize some critical features of the above evaluations, which will inspire a symmetric formula of double-trace amplitude with the $(g_i^-,g_j^-)$ configuration in the coming section:
\begin{itemize}
\item (i). There is a one-to-one correspondence between the refined graphs which contain  positive-helicity gravitons and gluons as nodes, and the spanning forests in four dimensions.

\item (ii). Although, gravitons are involved in the color-ordered YM amplitudes in the expansion (\ref{Eq:PureYMExpansion}), they do not appear in the Parke-Taylor factor (except the numerator $\Spaa{h_i,g_j}^4$ for the $(h^-_i,g^-_j)$ configuration) in the spanning forest formula (\ref{Eq:SingleTraceSpanningTree}). Thus the cyclic symmetry of gluons in \eqref{Eq:SingleTraceSpanningTree} is manifest.

\item (iii). The spanning forest formula is independent of the choice of gauge $\xi$ and $\eta$.
\end{itemize}
In the coming section, we generalize the spanning forest formula (\ref{Eq:SingleTraceSpanningTree}) to double-trace MHV amplitudes with the $(g_i^-,g_j^-)$ configuration.

\section{Double-trace MHV amplitudes with $(g^-_i,g^-_j)$ configuration }\label{sec:DoubleTraceConf1}
%
\begin{figure}
\centering
\includegraphics[width=0.6\textwidth]{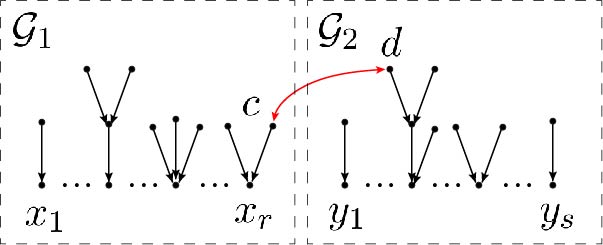}
\caption{ A typical spanning forest for double-trace amplitude with the $(g^-_i,g^-_j)$ configuration}\label{Fig:SpaningForestDouble1}
\end{figure}
\begin{figure}
\centering
\includegraphics[width=0.45\textwidth]{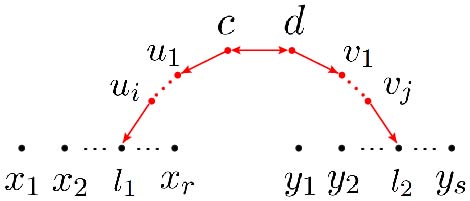}
\caption{ A bridge between the two traces in a spanning forest for a double-trace amplitude with the $(g^-_i,g^-_j)$ configuration }\label{Fig:Bridge}
\end{figure}
In this section, we generalize the pattern of single-trace MHV amplitudes (\ref{Eq:SingleTraceSpanningTree}) to double-trace ones with the  $(g^-_i,g^-_j)$ configuration. We prove that the double-trace MHV amplitude
 $A^{(g_i,g_j)}(x_1,\ldots, x_r|y_1,\ldots,y_s\Vert \mathsf{H})$  (where the two gluon traces are $\pmb{1}=\{x_1,\dots,x_r\}$, $\pmb{2}=\{y_1,\dots,y_s\}$ and the graviton set is $\mathsf{H}$) satisfies the following spanning forest formula
\bea
\boxed{\text{\begin{tabular}{c}
  $A^{(g_i,g_j)}(x_1,\ldots, x_r|y_1,\ldots,y_s\Vert \mathsf{H})~~~~~~~~~~~~~~~~~~~~~~~~~~~~~~~~~~~~~~~~~~~~~~~~~~~~~~~~~~~~~~~~~~~~~~~~$\\
  \\
  $=\frac{\Spaa{g_i,g_j}^4}{\left(x_1,\dots,x_r\right)\left(y_1,\dots,y_s\right)}\Bigg[\,\Sl_{\mathcal{G}=\mathcal{G}_1\oplus\mathcal{G}_2}\,\Sl_{\substack{c\in \mathcal{G}_1\\d\in \mathcal{G}_2}}\left(-k_c\cdot k_d\right)\frac{\Spaa{c,\zeta}\Spaa{d,\chi}}{\Spaa{c,d}\Spaa{\zeta,\chi}} \prod\limits_{{e(x,y)\in\,\mathcal{E}(\mathcal{G})}}\frac{\Spaa{\xi,y}\Spaa{\lambda_e,y}\Spbb{y,x}}{\Spaa{\xi,x}\Spaa{\lambda_e,x}\Spaa{y,x}}\Bigg]$  \\
\end{tabular}}}\,.\Label{Eq:DoubleTraceMHVgen1}
\eea
In the above expression, all possible spanning forests $\mathcal{G}$ with the pattern \figref{Fig:SpaningForestDouble1} are summed over. In each forest $\mathcal{G}$, the node set is given by $\pmb{1}\cup\pmb{2}\cup \mathsf{H}$ and all gluons in $\pmb{1}\cup\pmb{2}$ are considered as roots. Edges in a forest  $\mathcal{G}$ are presented by arrow lines pointing towards roots. Each forest $\mathcal{G}$ is composed of two sub-forests $\mathcal{G}_1$ and $\mathcal{G}_2$ whose roots live in $\pmb{1}$ and $\pmb{2}$ correspondingly. We further sum over all possible choices of  $c\in \mathcal{G}_1$ and $d\in \mathcal{G}_2$. For a given $\mathcal{G}=\mathcal{G}_1\oplus\mathcal{G}_2$ and a given choice of $c\in \mathcal{G}_1$ and $d\in \mathcal{G}_2$, a crossing factor
\bea
\left(-k_c\cdot k_d\right)\frac{\Spaa{c,\zeta}\Spaa{d,\chi}}{\Spaa{c,d}\Spaa{\zeta,\chi}} \Label{Eq:CrossingFactor}
 \eea
 is implied where $\zeta$ and $\chi$ are two reference spinors which respectively reflect the cyclic symmetries of the traces $\pmb{1}$ and $\pmb{2}$. For convenience, we express this crossing factor by connecting $c$ and $d$ via a double-arrow line, as shown by \figref{Fig:SpaningForestDouble1}. Any other edge in the graph is either an \emph{inner edge} or an \emph{outer edge}, which depends on whether it lives on the bridge (see \figref{Fig:Bridge}) between traces $\pmb{1}$ and $\pmb{2}$ or not. Each edge $e(x,y)$ where the arrow points from $x$ to $y$ is associated with a factor
\bea
\frac{\Spaa{\xi,y}\Spaa{\lambda_e,y}\Spbb{y,x}}{\Spaa{\xi,x}\Spaa{\lambda_e,x}\Spaa{y,x}},
\eea
where $\xi$ is the common reference spinor of all $e(x,y)$ which reflects the choice of gauge of the half-polarizations in the refined graphs. The $\lambda_e$  for each edge $e(x,y)$ is a reference spinor reflecting the cyclic symmetries of the gluon traces. In particular, it is defined by
\bea
\lambda_e=\Bigg\{
            \begin{array}{cc}
              \eta & \text{(if $e(x,y)$ is an outer edge )} \\
              \zeta &~~~~\text{(if $e(x,y)$ is an inner edge and $e(x,y)\in \mathcal{G}_1$)} \\
              \chi &~~~~\text{(if $e(x,y)$ is an inner edge and $e(x,y)\in \mathcal{G}_2$)} \\
            \end{array},\Label{Eq:Lambda}
\eea
where the reference spinors $\zeta$ and $\chi$ were already introduced before, while $\eta$ is a new reference spinor. \emph{It is worth pointing that the summations in the expression (\ref{Eq:DoubleTraceMHVgen1}) can  be rearranged by (i). first summing over all splits of the graviton set $\mathsf{H}\to \mathsf{H}_A, \mathsf{H}_B$; (ii). for a given split, summing over  all possible bridges $\mathcal{B}\left(\mathsf{H}_A\right)$, which involve all the gravitons in  $\mathsf{H}_A$ as nodes, between the two traces $\pmb{1}$ and $\pmb{2}$, (iii). for a given split and a given bridge, summing over all possible spanning forests $\mathcal{G}'$, in which all gravitons and gluons are considered as nodes and all elements in $\pmb{1}\cup\pmb{2}\cup\mathsf{H}_A$ are considered as roots.} The formula (\ref{Eq:DoubleTraceMHVgen1}) is then rewritten as an equivalent formula
\bea
\boxed{\text{\begin{tabular}{c}
  $A^{(g^-_i,g^-_j)}(x_1,\ldots, x_r|y_1,\ldots,y_s\Vert \mathsf{H})~~~~~~~~~~~~~~~~~~~~~~~~~~~~~~~~~~~~~~~~~~~~~~~~~~~~~~~~~~~~~~~~~~~~~~~~$\\
  \\
  $=\frac{\Spaa{g_i,g_j}^4}{\left(x_1,\dots,x_r\right)\left(y_1,\dots,y_s\right)}\Bigg[\,\Sl_{\mathsf{H}\to \mathsf{H}_A, \mathsf{H}_B}\,\Sl_{\mathcal{B}\,\left(\mathsf{H}_A\right)}\left(-k_c\cdot k_d\right)\frac{\Spaa{c,\zeta}\Spaa{d,\chi}}{\Spaa{c,d}\Spaa{\zeta,\chi}}\,\prod\limits_{{e(x,y)\in\,\mathcal{E}_1(\mathcal{B}(\mathsf{H}_A))}}\frac{\Spaa{\xi,y}\Spaa{\zeta,y}\Spbb{y,x}}{\Spaa{\xi,x}\Spaa{\zeta,x}\Spaa{y,x}}$\\
  $~~~~~~~~~~\times\prod\limits_{{e(x,y)\in\,\mathcal{E}_2(\mathcal{B}(\mathsf{H}_A))}}\frac{\Spaa{\xi,y}\Spaa{\chi,y}\Spbb{y,x}}{\Spaa{\xi,x}\Spaa{\chi,x}\Spaa{y,x}}\,
  \Sl_{\mathcal{G}'}\,\prod\limits_{{e(x,y)\in\,\mathcal{E}(\mathcal{G}')}}\frac{\Spaa{\xi,y}\Spaa{\eta,y}\Spbb{y,x}}{\Spaa{\xi,x}\Spaa{\eta,x}\Spaa{y,x}}\Bigg]$  \\
\end{tabular}}}\, ,\nn
\Label{Eq:DoubleTraceMHVgen2}
\eea
where $c$ and $d$ are the two gravitons connected by the double arrow line on the bridge $\mathcal{B}\left(\mathsf{H}_A\right)$. The $\mathcal{E}_1(\mathcal{B}(\mathsf{H}_A))$ and $\mathcal{E}_2(\mathcal{B}(\mathsf{H}_A))$ are the set of single-arrowed edges whose arrows points towards the traces $\pmb{1}$ and $\pmb{2}$, respectively. Graphs $\mathcal{G}'$ denote those spanning forests rooted at elements in $\pmb{1}\cup\pmb{2}\cup \mathsf{H}_A$. Thus all edges of $\mathcal{G}'$  are apparently those outer edges in \eqref{Eq:DoubleTraceMHVgen1} and \eqref{Eq:Lambda}.

In the following, we first investigate the symmetries of \eqref{Eq:DoubleTraceMHVgen1}. After that, we provide an example for  \eqref{Eq:DoubleTraceMHVgen1} and then the general proof by the refined graphic rule.

\subsection{Symmetries of the formula}\label{sec:SYM1}

The spanning forest formula  (\ref{Eq:DoubleTraceMHVgen1}) is much more symmetric than the expansion  (\ref{Eq:DoubleMHVYMExpansion1}) given by refined graphic rule.

First,  permutations in the Parke-Taylor factors of (\ref{Eq:DoubleMHVYMExpansion1}) involve all external gluons and gravitons. On the contrary, gravitons and gluon traces in  (\ref{Eq:DoubleTraceMHVgen1}) are disentangled from one another: Each trace is expressed by a Parke-Taylor factor with its own gluons in the original permutation, while gravitons and the other trace do not participate in. Therefore, \eqref{Eq:DoubleTraceMHVgen1} is explicitly symmetric under the exchanging of the two gluon traces. Moreover,  the invariance of \eqref{Eq:DoubleTraceMHVgen1} under exchanging any two gravitons seems more transparent because gravitons are already extracted out from the Parke-Taylor factors and only involved in the summation over spanning forests.

Second, there are ``gauge symmetries" corresponding to the arbitrariness of the reference spinors $\xi$, $\zeta$, $\chi$ and $\eta$. In the coming subsections, we will see $\xi$ comes from the reference spinor of all `half' polarizations, thus its arbitrariness is essentially the gauge symmetry of amplitudes. Nevertheless, the arbitrariness of the choices of $\lambda_e$ (i.e., $\zeta$, $\chi$ and $\eta$) that encode the cyclic symmetries of the two traces is not so clear. Now let us understand these symmetries of the expression (\ref{Eq:DoubleTraceMHVgen1}).

{\bf(i).} {\emph{The invariance of (\ref{Eq:DoubleTraceMHVgen1}) under the change of $\eta$}}~~This symmetry is easily understood when the amplitude is expressed by \eqref{Eq:DoubleTraceMHVgen2}: For a given split $\mathsf{H}\to \mathsf{H}_A, \mathsf{H}_B$ and a  given bridge between the two traces, the summation over all spanning forests rooted at elements in $\pmb{1}\cup\pmb{2}\cup\mathsf{H}_A$ exactly produces the determinant of the Hodges matrix (\ref{Eq:HodgesMatrix})  \cite{Feng:2012sy,Du:2016wkt}
\bea
\Sl_{\mathcal{G}'}\prod\limits_{{e(x,y)\in\,\mathcal{E}(\mathcal{G}')}}\frac{\Spaa{\xi,y}\Spaa{\eta,y}\Spbb{y,x}}{\Spaa{\xi,x}\Spaa{\eta,x}\Spaa{y,x}}=\det \left[\phi_{\mathsf{H}_B}\right],
\eea
which is independent of the choice of $\eta$.  Since each bridge is associated by a Hodges determinant that is independent of  $\eta$, $\eta$ can even be chosen differently for distinct configurations of bridge. Nevertheless, in this paper, we choose all $\eta$'s corresponding to different bridges as the same one for convenience.

{\bf(ii).} {\emph{The invariance of (\ref{Eq:DoubleTraceMHVgen1}) under the change of $\zeta$}}~~ Since $\eta$ can be chosen arbitrarily, we just set $\eta=\chi$, which does not bring any divergency to \eqref{Eq:DoubleTraceMHVgen1}. Now we proceed our discussion by classifying the terms inside the square brackets  of \eqref{Eq:DoubleTraceMHVgen1} according to \emph{whether $c\in \mathcal{G}_1$ is a gluon or a graviton} (i.e. $c\in\pmb{1}$ or $c\in \mathcal{G}_1\setminus\pmb{1}$), for  given $\mathcal{G}_1$ and $\mathcal{G}_2$. In the latter case, there exist single-arrowed edges pointing towards $\pmb{1}$ on the bridge between $\pmb{1}$ and $\pmb{2}$ (see \figref{Fig:Bridge}), while in the former there is no such line on the bridge.  The contribution of terms with $c\in \pmb{1}$ is given by
\bea
T_a=\Sl_{\substack{c\in \pmb{1}\\d\in \mathcal{G}_2}}\left(-k_c\cdot k_d\right)\frac{\Spaa{c,\zeta}\Spaa{d,\chi}}{\Spaa{c,d}\Spaa{\zeta,\chi}} \Biggl[\,\prod\limits_{{e(x,y)\in\,\mathcal{E}\left(\mathcal{G}=\mathcal{G}_1\oplus\mathcal{G}_2\right)}}\frac{\Spaa{\xi,y}\Spaa{\chi,y}\Spbb{y,x}}{\Spaa{\xi,x}\Spaa{\chi,x}\Spaa{y,x}}\Biggr],
\eea
and the difference between $T'_a$ and $T_a$ corresponding to $\zeta'$ and $\zeta$ is evaluated as
\bea
\Delta T_a=\Sl_{\substack{c\in \pmb{1}\\d\in \mathcal{G}_2}}\frac{1}{2}\Spbb{c,d}\Spaa{d,\chi}\Spaa{\chi,c}\frac{\Spaa{\zeta',\zeta}}{\Spaa{\zeta',\chi}\Spaa{\zeta,\chi}} \Biggl[\,\prod\limits_{{e(x,y)\in\,\mathcal{E}\left(\mathcal{G}=\mathcal{G}_1\oplus\mathcal{G}_2\right)}}\frac{\Spaa{\xi,y}\Spaa{\chi,y}\Spbb{y,x}}{\Spaa{\xi,x}\Spaa{\chi,x}\Spaa{y,x}}\Biggr],
\eea
where Schouten identity (\ref{Eq:SpinorProp2}) has been applied. On another hand, the expression inside the square brackets  of \eqref{Eq:DoubleTraceMHVgen1} when $c\in \mathcal{G}_1\setminus\pmb{1}$ reads
\bea
T_b&=&\Sl_{\substack{c\in \mathcal{G}_1\setminus\pmb{1}\\d\in \mathcal{G}_2}}\left(-k_c\cdot k_d\right)\frac{\Spaa{c,\zeta}\Spaa{d,\chi}}{\Spaa{c,d}\Spaa{\zeta,\chi}} \nn
&\times&\prod\limits_{e(x,y){\in}\,\mathcal{E}_1(c) }\frac{\Spaa{\xi,y}\Spaa{\zeta,y}\Spbb{y,x}}{\Spaa{\xi,x}\Spaa{\zeta,x}\Spaa{y,x}}
\prod\limits_{\substack{e(x,y)\in\,\mathcal{E}(\mathcal{G}_1)\\e(x,y)\cancel{\in}\,\mathcal{E}_1(c) }}\frac{\Spaa{\xi,y}\Spaa{\chi,y}\Spbb{y,x}}{\Spaa{\xi,x}\Spaa{\chi,x}\Spaa{y,x}} \prod\limits_{{e(x,y)\in\,\mathcal{E}(\mathcal{G}_2)}}\frac{\Spaa{\xi,y}\Spaa{\chi,y}\Spbb{y,x}}{\Spaa{\xi,x}\Spaa{\chi,x}\Spaa{y,x}},\Label{Eq:Gauge1}
\eea
where $\mathcal{E}_1(c)$ denotes the set of \emph{single-arrowed inner edges} that are pointing towards the trace $\pmb{1}$ for a given $c$.
Supposing the bridge between the two traces $\pmb{1}$ and $\pmb{2}$ is given by \figref{Fig:Bridge},
 the first product in \eqref{Eq:Gauge1}  for a given $c\in \mathcal{G}_1\setminus \pmb{1}$ and $d\in \mathcal{G}_2$  reads
\bea
&&\prod\limits_{e(x,y){\in}\,\mathcal{E}_1(c) }\frac{\Spaa{\xi,y}\Spaa{\zeta,y}\Spbb{y,x}}{\Spaa{\xi,x}\Spaa{\zeta,x}\Spaa{y,x}}\nn
&=&\frac{\Spaa{\xi,u_1}\cancel{\Spaa{\zeta,u_1}}\Spbb{u_1,c}}{\Spaa{\xi,c}\Spaa{\zeta,c}\Spaa{u_1,c}}\,\frac{\Spaa{\xi,u_2}\cancel{\Spaa{\zeta,u_2}}\Spbb{u_2,u_1}}{\Spaa{\xi,u_1}\cancel{\Spaa{\zeta,u_1}}\Spaa{u_2,u_1}}\, \dots \,\frac{\Spaa{\xi,l_1}\Spaa{\zeta,l_1}\Spbb{l_1,u_i}}{\Spaa{\xi,u_i}\cancel{\Spaa{\zeta,u_i}}\Spaa{l_1,u_i}}\nn
&=&\Biggl[\frac{\Spaa{\xi,u_1}\Spaa{\chi,u_1}\Spbb{u_1,c}}{\Spaa{\xi,c}\Spaa{\chi,c}\Spaa{u_1,c}}\,\frac{\Spaa{\xi,u_2}\Spaa{\chi,u_2}\Spbb{u_2,u_1}}{\Spaa{\xi,u_1}\Spaa{\chi,u_1}\Spaa{u_2,u_1}}\, \dots \,\frac{\Spaa{\xi,l_1}\Spaa{\chi,l_1}\Spbb{l_1,u_i}}{\Spaa{\xi,u_i}\Spaa{\chi,u_i}\Spaa{l_1,u_i}}\Biggr]\,\frac{\Spaa{\chi,c}}{\Spaa{\chi,l_1}}\frac{\Spaa{\zeta,l_1}}{\Spaa{\zeta,c}}\nn
&=&\Biggl[\,\prod\limits_{e(x,y){\in}\,\mathcal{E}_1(c) }\frac{\Spaa{\xi,y}\Spaa{\chi,y}\Spbb{y,x}}{\Spaa{\xi,x}\Spaa{\chi,x}\Spaa{y,x}}\Biggr]\,\frac{\Spaa{\chi,c}}{\Spaa{\chi,l_1}}\frac{\Spaa{\zeta,l_1}}{\Spaa{\zeta,c}}.\Label{Eq:InnerEdges1}
\eea
The expression (\ref{Eq:Gauge1}) is therefore arranged as
\bea
T_b&=&\Biggl[\Sl_{\substack{c\in \mathcal{G}_1\setminus{\pmb{1}}\\d\in \mathcal{G}_2}}\frac{1}{2}\Spbb{c,d}\Spaa{\chi,d}\frac{\Spaa{\chi,c}}{\Spaa{\chi,l_1}}\frac{\Spaa{\zeta,l_1}}{\Spaa{\zeta,\chi}}\Biggr] \Biggl[\,\prod\limits_{{e(x,y)\in\,\mathcal{E}\left(\mathcal{G}=\mathcal{G}_1\oplus\mathcal{G}_2\right)}}\frac{\Spaa{\xi,y}\Spaa{\chi,y}\Spbb{y,x}}{\Spaa{\xi,x}\Spaa{\chi,x}\Spaa{y,x}}\Biggr].\Label{Eq:Gauge2}
\eea
\begin{figure}
\centering
\includegraphics[width=0.8\textwidth]{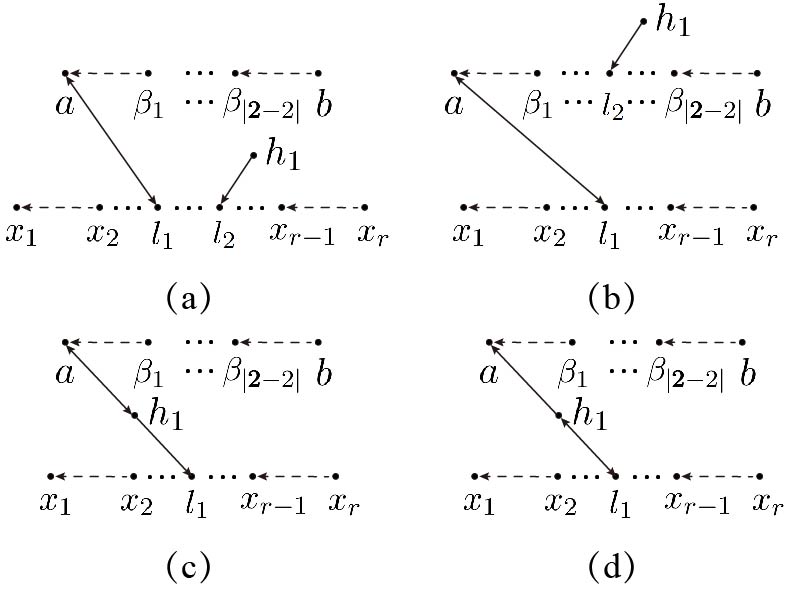}
\caption{Typical refined graphs for the double-trace amplitude $A_{m=2}^{\text{MHV}}(x_1,\ldots, x_r|y_1,\ldots,y_s\Vert h_1)$}\label{Fig:DoubleTraceExample1}
\end{figure}
Apparently, only the first factor involves $\zeta$. When $\zeta$ is replaced by $\zeta'$, the first factor differs from the original one (the one with $\zeta$) by
\bea
\Sl_{\substack{c\in \mathcal{G}_1\setminus{\pmb{1}}\\d\in \mathcal{G}_2}}\frac{1}{2}\Spbb{c,d}\Spaa{\chi,d}\frac{\Spaa{\chi,c}}{\Spaa{\chi,l_1}}\left(\frac{\Spaa{\zeta',l_1}}{\Spaa{\zeta',\chi}}-\frac{\Spaa{\zeta,l_1}}{\Spaa{\zeta,\chi}}\right)
&=&\Biggl[\Sl_{\substack{c\in \mathcal{G}_1\setminus\pmb{1}\\d\in \mathcal{G}_2}}\frac{1}{2}\Spbb{c,d}\Spaa{\chi,d}\Spaa{\chi,c}\Biggr]\frac{\Spaa{\zeta',\zeta}}{\Spaa{\chi,\zeta'}\Spaa{\chi,\zeta}}
\eea
where we have applied Schouten identity (\ref{Eq:SpinorProp2}). Thus
\bea
\Delta T_b=\Sl_{\substack{c\in \mathcal{G}_1\setminus\pmb{1}\\d\in \mathcal{G}_2}}\frac{1}{2}\Spbb{c,d}\Spaa{d,\chi}\Spaa{\chi,c}\frac{\Spaa{\zeta',\zeta}}{\Spaa{\chi,\zeta'}\Spaa{\chi,\zeta}}\Biggl[\,\prod\limits_{{e(x,y)\in\,\mathcal{E}\left(\mathcal{G}=\mathcal{G}_1\oplus\mathcal{G}_2\right)}}\frac{\Spaa{\xi,y}\Spaa{\chi,y}\Spbb{y,x}}{\Spaa{\xi,x}\Spaa{\chi,x}\Spaa{y,x}}\Biggr].
\eea
The sum of $\Delta T_a$ and $\Delta T_b$ is then given by
\bea
\Delta T_a+\Delta T_b&=&\Sl_{\substack{c\in \mathcal{G}_1\\d\in \mathcal{G}_2}}\frac{1}{2}\Spbb{c,d}\Spaa{d,\chi}\Spaa{\chi,c}\frac{\Spaa{\zeta',\zeta}}{\Spaa{\chi,\zeta'}\Spaa{\chi,\zeta}}\Biggl[\,\prod\limits_{{e(x,y)\in\,\mathcal{E}\left(\mathcal{G}=\mathcal{G}_1\oplus\mathcal{G}_2\right)}}\frac{\Spaa{\xi,y}\Spaa{\chi,y}\Spbb{y,x}}{\Spaa{\xi,x}\Spaa{\chi,x}\Spaa{y,x}}\Biggr]\nn
&=&-\Sl_{c,d\in \mathcal{G}_1}\frac{1}{2}\Spbb{c,d}\Spaa{d,\chi}\Spaa{\chi,c}\frac{\Spaa{\zeta',\zeta}}{\Spaa{\chi,\zeta'}\Spaa{\chi,\zeta}}\Biggl[\,\prod\limits_{{e(x,y)\in\,\mathcal{E}\left(\mathcal{G}=\mathcal{G}_1\oplus\mathcal{G}_2\right)}}\frac{\Spaa{\xi,y}\Spaa{\chi,y}\Spbb{y,x}}{\Spaa{\xi,x}\Spaa{\chi,x}\Spaa{y,x}}\Biggr],
\eea
where momentum conservation has been applied. For  given $c,d \in \mathcal{G}_1$, the last line of the above equation is antisymmetric about $c$ and $d$, thus has to vanish when  all nodes $c,d \in \mathcal{G}_1$ are summed over.
We then conclude that \eqref{Eq:DoubleTraceMHVgen2} (thus \eqref{Eq:DoubleTraceMHVgen1}) is invariant under the change of $\zeta$.

{\bf(iii).} {\emph{The invariance of (\ref{Eq:DoubleTraceMHVgen1}) under the change of $\chi$}}~~When we choose $\eta=\zeta$, this symmetry naturally follows from a discussion  parallel with (ii) because
 (\ref{Eq:DoubleTraceMHVgen1}) has a symmetric form under the exchange of the roles of $\pmb{1}$ and $\pmb{2}$.

\subsection{An example study}

Before showing the general proof of the formula (\ref{Eq:DoubleTraceMHVgen1}), we explicitly evaluate the double-trace amplitude $A^{(g_i^-,g_j^-)}(x_1,\ldots, x_r|y_1,\ldots,y_s\Vert h_1)$ involving only one graviton. According to the refined graphic rule given in \secref{sec:ExpansionDoubleTrace}, this amplitude can be written as \eqref{Eq:DoubleMHVYMExpansion1}, in which all typical refined graphs $\mathcal{F}\in \mathcal{F}[0,2]$ are presented by \figref{Fig:DoubleTraceExample1}.

We first evaluate the graphs with the structures \figref{Fig:DoubleTraceExample1} (a) and (b), where the graviton $h_1$ plays as an outer node. The contribution of all graphs with the structure \figref{Fig:DoubleTraceExample1} (a) is given by
\bea
T_1&=&\Sl_{\pmb{\gamma}}\,\Sl_{\pmb{\sigma}\,\in\,\pmb{\gamma}\,\shuffle\,\{h_1\}}(-k_a\cdot k_{l_1})\left(\epsilon_{h_1}\cdot k_{l_2}\right)\,A\left(x_1,\pmb{\sigma},x_r\right)\nn
&=&\Sl_{\pmb{\gamma}}(-k_a\cdot k_{l_1})\,\Sl_{\pmb{\sigma}\,\in\,\pmb{\gamma}\,\shuffle\,\{h_1\}}\frac{\Spaa{\xi,l_2}\Spbb{l_2,h_1}}{\Spaa{\xi,h_1}}\,\frac{\Spaa{g_i,g_j}^4}{\left(x_1,\pmb{\sigma},x_r\right)}\nn
&=&\Sl_{\pmb{\gamma}}(-k_a\cdot k_{l_1})\,\frac{\Spaa{g_i,g_j}^4}{\left(x_1,\pmb{\gamma},x_r\right)}\,\frac{\Spaa{\xi,l_2}\Spaa{x_r, l_2}\Spbb{l_2,h_1}}{\Spaa{\xi,h_1}\Spaa{x_r,h_1}\Spaa{l_2,h_1}}.\Label{Eq:DoubleTraceExample1}
\eea
where $l_2\in\pmb{1}\setminus\{x_r\}=\{x_1,\dots,x_{r-1}\}$ and the permutations $\pmb{\gamma}$ obeys
\bea
\pmb{\gamma}\in\left\{x_2,\dots,l_1=x_a,\{x_{a+1},\dots,x_{r-1}\}\shuffle\{a,\beta_1,\dots,\beta_{|\pmb{2}|-2},b\}\right\}.
\eea
On the second line of \eqref{Eq:DoubleTraceExample1}, we have written $\epsilon_{h_1}(\xi)\cdot k_{l_2}$ by spinor products and the MHV Yang-Mills amplitudes $A\left(x_1,\pmb{\sigma},x_r\right)$
by Parke-Taylor formula. On the third line of \eqref{Eq:DoubleTraceExample1}, the identity (\ref{Eq:PTRelation1}) was applied. Following a similar discussion, the contribution of \figref{Fig:DoubleTraceExample1} (b) reads
\bea
T_2&=&\Sl_{\pmb{\gamma}}(-k_a\cdot k_{l_1})\,\frac{\Spaa{g_i,g_j}^4}{\left(x_1,\pmb{\gamma},x_r\right)}\,\frac{\Spaa{\xi,l_2}\Spaa{x_r, l_2}\Spbb{l_2,h_1}}{\Spaa{\xi,h_1}\Spaa{x_r,h_1}\Spaa{l_2,h_1}}.~~(l_2\in\pmb{2}).
\eea
When summed over all possible $l_2$ in $T_1$ and $T_2$, we arrive the total contribution of graphs with structures \figref{Fig:DoubleTraceExample1} (a) and (b):
\bea
I_1=\Sl_{l_2\in\pmb{1}\setminus\{x_r\}}T_1+\Sl_{l_2\in\pmb{2}}T_2=\left[\Sl_{\pmb{\gamma}}(-k_a\cdot k_{l_1})\,\frac{\Spaa{g_i,g_j}^4}{\left(x_1,\pmb{\gamma},x_r\right)}\right]\,\left[\Sl_{l_2\in\pmb{1}\cup\pmb{2}}\frac{\Spaa{\xi,l_2}\Spaa{x_r,l_2}\Spbb{l_2,h_1}}{\Spaa{\xi,h_1}\Spaa{x_r,h_1}\Spaa{l_2,h_1}}\right],
\eea
where the case $l_2=x_r$ is already involved in the second equality since $\Spaa{x_r,x_r}=0$. The first factor in the above expression is explicitly written as
\bea
&&\Sl_{\pmb{\gamma}}(-k_a\cdot k_{l_1})\frac{\Spaa{g_i,g_j}^4}{\left(x_1,\dots,l_1=x_a,\pmb{\gamma}\in \{x_{a+1},\dots,x_{r-1}\}\shuffle\{a,\pmb{\beta},b\},x_r\right)}\nn
&=& (-k_a\cdot k_{l_1})\frac{\Spaa{g_i,g_j}^4}{\left(x_1,\dots,x_r\right)}\,\frac{\Spaa{l_1,x_r}}{\left\langle l_1,a,\pmb{\beta},b,x_r \, \right\rangle }\nn
&=& \frac{\Spaa{g_i,g_j}^4}{\left(x_1,\dots,x_r\right)}\,\frac{1}{\left(a,\pmb{\beta},b\right) }(-k_a\cdot k_{l_1})\frac{\Spaa{b,a}\Spaa{l_1,x_r}}{\Spaa{l_1,a}\Spaa{b,x_r}},
\eea
in which, the identity (\ref{Eq:PTRelation2}) has been applied on the second line. When  all possible $\pmb{\beta}$ (with the sign $(-1)^{|\,\pmb{2}, a,b\,|}$) are summed over, the factor $\frac{1}{\left(a,\pmb{\beta},b\right) }$ turns into
$\frac{1}{\left(y_1,\dots,y_s\right)}$,
according to the Kleiss-Kuijf (KK) relation \cite{Kleiss:1988ne} (\ref{Eq:KK}) between Parke-Taylor factors.
Therefore all contributions  of graphs with the structures \figref{Fig:DoubleTraceExample1} (a) and (b)  are collected as
\bea
I_1=\frac{\Spaa{g_i,g_j}^4}{\left(x_1,\dots,x_r\right)\left(y_1,\dots,y_s\right)}\left[\,\Sl_{l_1\in\pmb{1},a\in\pmb{2}}(-k_a\cdot k_{l_1})\frac{\Spaa{a,b}\Spaa{l_1,x_r}}{\Spaa{l_1,a}\Spaa{x_r,b}}\right]\,\left[\Sl_{l_2\in\pmb{1}\cup\pmb{2}}\frac{\Spaa{\xi,l_2}\Spaa{x_r,l_2}\Spbb{l_2,h_1}}{\Spaa{\xi,h_1}\Spaa{x_r,h_1}\Spaa{l_2,h_1}}\right].
\eea

Each graph with the structure \figref{Fig:DoubleTraceExample1} (c) contributes
\bea
T_3&=&(-k_a\cdot k_{h_1})(\epsilon_{h_1}\cdot k_{l_1})\Sl_{\pmb{\sigma}}\frac{\Spaa{g_i,g_j}^4}{\left(x_1,x_2,\dots,l_1=x_a,\pmb{\sigma}\in\{x_{a+1},\dots,x_{r-1}\}\shuffle\{h_1,a,\pmb{\beta},b\},x_r\right)}.
\eea
Noting that  permutation $\pmb{\sigma}$ in each Parke-Taylor factor can be rewritten as follows
\bea
\pmb{\sigma}\in \{x_{a+1},\dots,x_{r-1}\}\shuffle\{h_1\}\shuffle\{a,\pmb{\beta},b\}\big|_{h_1\prec a},
\eea
we are able to apply the identity (\ref{Eq:PTRelation2}) to extract the trace $\pmb{2}$ and the graviton $h_1$ from the Parke-Taylor factor in turn. The $T_3$ is then expressed as
\bea
T_3&=&(-k_a\cdot k_{h_1})(\epsilon_{h_1}\cdot k_{l_1})\Sl_{\pmb{\gamma}}\frac{\Spaa{g_i,g_j}^4}{\left(x_1,\dots,l_1=x_a,\pmb{\gamma}\in\{x_{a+1},\dots,x_{r-1}\}\shuffle\{h_1\},x_r\right)}\frac{1}{\left( a,\pmb{\beta},b\right)}\frac{\Spaa{h_1,x_r}\Spaa{b,a}}{\Spaa{h_1,a}\Spaa{b,x_r}}\nn
&=&\frac{\Spaa{g_i,g_j}^4}{\left(x_1,\dots,x_r\right)}\frac{1}{\left( a,\pmb{\beta},b\right)}\left[(-k_a\cdot k_{h_1})\frac{\Spaa{h_1,x_r}\Spaa{b,a}}{\Spaa{h_1,a}\Spaa{b,x_r}}\right]\left[\frac{\Spaa{\xi,l_1}\Spaa{x_r, l_1}\Spbb{l_1,h_1}}{\Spaa{\xi,h_1}\Spaa{x_r,h_1}\Spaa{l_1,h_1}}\right].
\eea
When all possible $\pmb{\beta}$ (with the sign $(-1)^{|\,\pmb{2}, a,b\,|}$), $a\in \pmb{2}\setminus\{b\}$, $l_1\in\pmb{1}\setminus\{x_r\}$ are summed over and the KK relation \cite{Kleiss:1988ne} is applied to the trace $\pmb{2}$, the contribution of all graphs with the structure \figref{Fig:DoubleTraceExample1} (c) becomes
\bea
I_2=\frac{\Spaa{g_i,g_j}^4}{\left(x_1,\dots,x_r\right)\left(y_1,\dots,y_s\right)}\left[\,\Sl_{a\in\pmb{2}}(-k_a\cdot k_{h_1})\frac{\Spaa{h_1,x_r}\Spaa{b,a}}{\Spaa{h_1,a}\Spaa{b,x_r}}\,\right]\left[\,\Sl_{l_1\in\pmb{1}}\frac{\Spaa{\xi,l_1}\Spaa{x_r,l_1}\Spbb{l_1,h_1}}{\Spaa{\xi,h_1}\Spaa{x_r,h_1}\Spaa{l_1,h_1}}\,\right].
\eea
Here, the terms with $a=b$ and $l_1=x_r$ were also included in the corresponding summations since $\Spaa{b,b}=\Spaa{x_r,x_r}=0$.

 Each graph with the structure \figref{Fig:DoubleTraceExample1} (d) provides a contribution
\bea
T_4&=&(-k_{h_1}\cdot k_{l_1})(-\epsilon_{h_1}\cdot k_{a})\Sl_{\pmb{\gamma}}\frac{\Spaa{g_i,g_j}^4}{\left(x_1,x_2,\dots,l_1=x_a,\pmb{\gamma}\in\{x_{a+1},\dots,x_{r-1}\}\shuffle\{h_1,a,\pmb{\beta},b\},x_r\right)}.
\eea
When the identity (\ref{Eq:PTRelation2}) is applied, $T_4$ turns into
\bea
T_4&=&(-k_{h_1}\cdot k_{l_1})(-\epsilon_{h_1}\cdot k_{a}) \frac{\Spaa{g_i,g_j}^4}{\left(x_1,\dots,x_r\right)}\frac{\Spaa{l_1,x_r}}{\Spaa{l_1, h_1,a,\pmb{\beta},b, x_r}}\nn
&=&(-\epsilon_{h_1}\cdot k_{a}) \frac{\Spaa{g_i,g_j}^4}{\left(x_1,\dots,x_r\right)}\frac{1}{\left(h_1,a,\pmb{\beta},b\right)}\left[(-k_{h_1}\cdot k_{l_1})\frac{\Spaa{l_1,x_r}\Spaa{b,h_1}}{\Spaa{l_1,h_1}\Spaa{b,x_r}}\right],
\eea
where $\Spaa{a_1,a_2,\dots,a_i}\equiv \Spaa{a_1,a_2}\Spaa{a_2,a_3}\dots\Spaa{a_{i-1},a_i}$. We further apply the identity (\ref{Eq:PTRelation1}) to extract $h_1$ from the trace $\pmb{2}$ (i.e. the Parke-Taylor factor ${1}/{\left(h_1,a,\pmb{\beta},b\right)}$) and write $\epsilon_{h_1}\cdot k_{a}$ by spinor products, then obtain
\bea
T_4=\frac{\Spaa{g_i,g_j}^4}{\left(x_1,\dots,x_r\right)}\frac{1}{\left( a,\pmb{\beta},b\right)}\left[(-k_{h_1}\cdot k_{l_1})\frac{\Spaa{l_1,x_r}\Spaa{b,h_1}}{\Spaa{l_1,h_1}\Spaa{b,x_r}}\right]\left[\frac{\Spaa{\xi,a}\Spaa{b,a}\Spbb{a,h_1}}{\Spaa{\xi,h_1}\Spaa{b,h_1}\Spaa{a,h_1}}\right].
\eea
Summing over all $\pmb{\beta}\in \mathsf{KK}[\pmb{2},a,b]$ (with the sign $(-1)^{|\,\pmb{2}, a,b\,|}$), $a\in\pmb{2}\setminus\{b\}$, $l_1\in\pmb{1}\setminus\{x_r\}$ and applying the KK relation on the Parke-Taylor factors of trace $\pmb{2}$, we get
\bea
I_3=\frac{\Spaa{g_i,g_j}^4}{\left(x_1,\dots,x_r\right)\left( y_1,\dots,y_s\right)}\left[\Sl_{l_1\in\pmb{1}}(-k_{h_1}\cdot k_{l_1})\frac{\Spaa{l_1,x_r}\Spaa{b,h_1}}{\Spaa{l_1,h_1}\Spaa{b,x_r}}\right]\left[\Sl_{a\in\pmb{2}}\frac{\Spaa{\xi,a}\Spaa{b,a}\Spbb{a,h_1}}{\Spaa{\xi,h_1}\Spaa{b,h_1}\Spaa{a,h_1}}\right],
\eea
where $a=b$ and $l_1=x_r$ are included again due to the antisymmetry of the spinor products.
\begin{figure}
\centering
\includegraphics[width=1\textwidth]{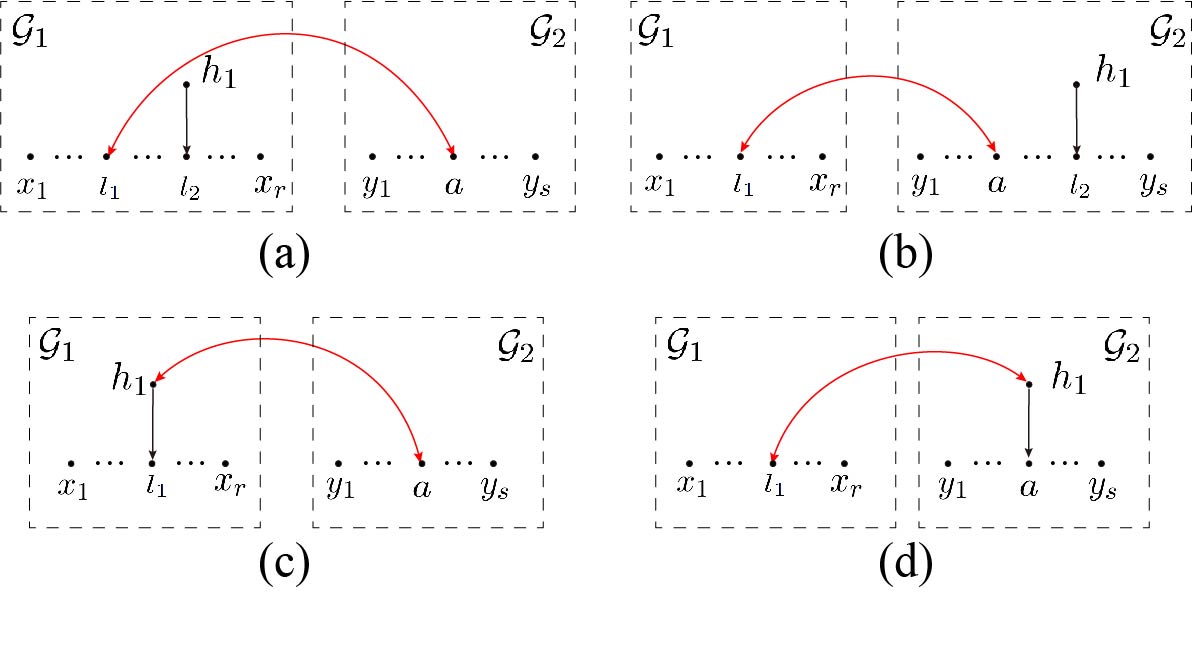}
\caption{Spanning forests for the double-trace amplitude $A^{(g_i,g_j)}(x_1,\ldots, x_r|y_1,\ldots,y_s\Vert h_1)$.}\label{Fig:DoubleTraceExample2}
\end{figure}

The contribution of all refined graphs with the structures \figref{Fig:DoubleTraceExample1} are obtained by summing $I_1$, $I_2$ and $I_3$ together:
\bea
&&A^{(g_i,g_j)}(x_1,\ldots, x_r|y_1,\ldots,y_s\Vert h_1)\nn
&=&\,\frac{\Spaa{g_i,g_j}^4}{\left(x_1,\dots,x_r\right)\left(y_1,\dots,y_s\right)}\nn
&&~~~\times\Biggl[\,\Sl_{l_1\in\pmb{1},a\in\pmb{2}}(-k_a\cdot k_{l_1})\frac{\Spaa{l_1,x_r}\Spaa{a,b}}{\Spaa{l_1,a}\Spaa{x_r,b}}\,\Sl_{l_2\in\pmb{1}\cup\pmb{2}}\frac{\Spaa{\xi,l_2}\Spaa{l_2,x_r}\Spbb{l_2,h_1}}{\Spaa{\xi,h_1}\Spaa{h_1,x_r}\Spaa{l_2,h_1}}\nn
&&~~~~~~~\,\,+\Sl_{a\in\pmb{2}}(-k_a\cdot k_{h_1})\frac{\Spaa{h_1,x_r}\Spaa{a,b}}{\Spaa{h_1,a}\Spaa{x_r,b}}\,\,\,\,\,\Sl_{l_1\in\pmb{1}}\frac{\Spaa{\xi,l_1}\Spaa{l_1,x_r}\Spbb{l_1,h_1}}{\Spaa{\xi,h_1}\Spaa{h_1,x_r}\Spaa{l_1,h_1}}\nn
&&~~~~~~~\,\,+\Sl_{l_1\in\pmb{1}}(-k_{h_1}\cdot k_{l_1})\frac{\Spaa{l_1,x_r}\Spaa{h_1,b}}{\Spaa{l_1,h_1}\Spaa{x_r,b}}\,\,\Sl_{a\in\pmb{2}}\frac{\Spaa{\xi,a}\Spaa{a,b}\Spbb{a,h_1}}{\Spaa{\xi,h_1}\Spaa{h_1,b}\Spaa{a,h_1}}\,\Biggr].\Label{Eq:DoubleTraceExample2}
\eea
which precisely reproduces the formula (\ref{Eq:DoubleTraceMHVgen1}) with the choice of gauge $\zeta=x_r$ and $\chi=b$. In this example, all possible graphs expressed by the spanning forests in \eqref{Eq:DoubleTraceMHVgen1} are given by \figref{Fig:DoubleTraceExample2}, where both \figref{Fig:DoubleTraceExample2} (a) and (b) contribute to the first term, \figref{Fig:DoubleTraceExample2} (c) and (d) correspond to the second and the third term.
\emph{The cyclic symmetry of the traces demand that the above expression must be preserved when we consider any other gluons $x_u\in\pmb{1}$ and $y_v\in\pmb{2}$ as the last ones instead of $x_r$ and $b$ respectively. This has already been guaranteed by the symmetry under choosing different $\zeta$ and $\chi$ in (\ref{Eq:DoubleTraceMHVgen1}).}

\subsection{The general proof}\label{sec:GenProof}

Inspired by the above example and the study of single-trace amplitudes, we now prove the general formula (\ref{Eq:DoubleTraceMHVgen1}) by three steps: {\bf step-1.} extracting all outer gravitons from the Parke-Taylor factors in \eqref{Eq:DoubleMHVYMExpansion1}, {\bf step-2.} separating the two traces, which may be attached by inner gravitons, from one another, {\bf step-3.} extracting the inner gravitons from the corresponding Parke-Taylor factor. The explicit proof is following.

{\bf Step-1:} For a given refined graph $\mathcal{F}$  in \eqref{Eq:DoubleMHVYMExpansion1},  we pick a leaf (i.e. an outermost graviton) $h_a$. Permutations
established by $\mathcal{F}$ can be written as $\{x_1,\left(\mathcal{F}|_{x_1}\setminus\{x_1,h_a,x_r\}\right)\shuffle\{h_a\}|_{l\prec h_a},x_r\}$, where $l$ is supposed to be the node adjacent to $h_a$.
The coefficient corresponding to the graph $\mathcal{F}$ then reads
\bea
\mathcal{C}^{\,\mathcal{F}}=\mathcal{C}^{\,\mathcal{F}\setminus\{h_a\}}\,\left(\epsilon_{h_a}\cdot k_l\right).
\eea
\begin{figure}
\centering
\includegraphics[width=1\textwidth]{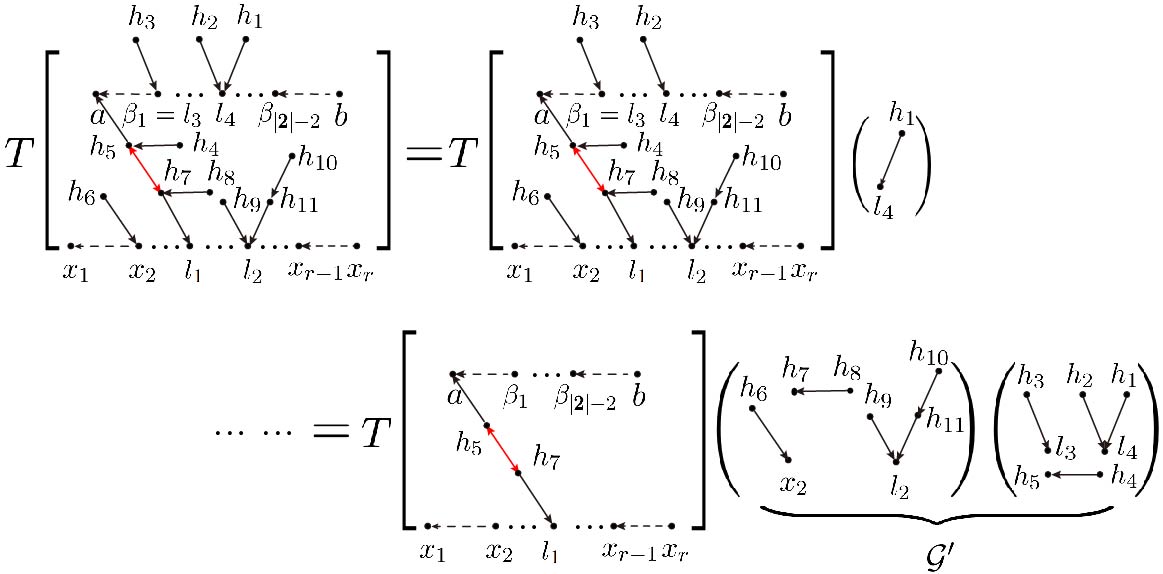}
\caption{An example for step-1 in the general proof of the  formula (\ref{Eq:DoubleTraceMHVgen1}). Here, the permutation $\pmb{\beta}$ satisfies $\pmb{\beta}\in \mathsf{KK}[\pmb{2},a,b]$. The graviton sets $\{h_1,h_2,h_3\}$, $\{h_4,h_8\}$ and $\{h_6, h_9,h_{10}, h_{11}\}$ are the sets of outer gravitons which belong to those trees planted at gluons in the trace $\pmb{2}$ (i.e. $l_3$ and $l_4$), gluons in the trace $\pmb{1}$ (i.e. $l_1$ and $l_2$) and the inner gravitons (i.e. $h_5$ and $h_7$), respectively. After manipulations in this step, only the two gluon traces as well as the inner gravitons $h_5$, $h_7$ remain in the refined graph. All other gravitons (and the tree structures associated to them) have been extracted out and described by the spanning forest $\mathcal{G}'$. }\label{Fig:DoubleTraceRecursion}
\end{figure}
The YM amplitudes   in \eqref{Eq:DoubleMHVYMExpansion1} corresponding to the graph $\mathcal{F}$ are collected as
\bea
\Sl_{\pmb{\sigma}\in\mathcal{F}|_{x_1}\setminus\{x_1, x_r\}}\,\frac{\Spaa{g_i,g_j}^4}{\left( x_1,\pmb{\sigma},x_r\right)}&=&\Sl_{\pmb{\gamma}\in \mathcal{F}|_{x_1}\setminus\{x_1,h_a,x_r\}}\,\Sl_{\pmb{\sigma}\in (\pmb{\gamma}\shuffle\{h_a\})|_{l\prec h_a}}\,\frac{\Spaa{g_i,g_j}^4}{\left( x_1,\pmb{\sigma},x_r\right)}\nn
&=&\Sl_{\pmb{\gamma}\in \mathcal{F}|_{x_1}\setminus\{x_1,h_a,x_r\}}\,\frac{\Spaa{g_i,g_j}^4}{\left( x_1,\pmb{\gamma},x_r\right)}\,\frac{\Spaa{l,x_r}}{\Spaa{l,h_a}\Spaa{h_a,x_r}},
\eea
where the identity (\ref{Eq:PTRelation1}) has been applied. Hence the total contribution of the graph $\mathcal{F}$ is recursively expressed  as
\bea
T^{\mathcal{F}}&=&\Biggl[\mathcal{C}^{\,\mathcal{F}\setminus\{h_a\}}\Sl_{\pmb{\gamma}\in \mathcal{F}|_{x_1}\setminus\{x_1,h_a,x_r\}}\,\frac{\Spaa{g_i,g_j}^4}{\left( x_1,\pmb{\gamma},x_r\right)}\,\Biggr]\Biggl[(\epsilon_{h_a}\cdot k_l)\,\frac{\Spaa{l,x_r}}{\Spaa{l,h_a}\Spaa{h_a,x_r}}\Biggr]\nn
&=&\Bigg[\,\mathcal{C}^{\,\mathcal{F}\setminus\{h_a\}}\Sl_{\pmb{\gamma}\in \mathcal{F}|_{x_1}\setminus\{x_1,h_a,x_r\}}\,\frac{\Spaa{g_i,g_j}^4}{\left( x_1,\pmb{\gamma},x_r\right)}\Bigg]\,\frac{\Spaa{\xi,l}\Spaa{x_r,l}\Spbb{l,h_a}}{\Spaa{\xi,h_a}\Spaa{x_r,h_a}\Spaa{l,h_a}}\nn
&=&T^{\mathcal{F}\setminus\{h_a\}}\frac{\Spaa{\xi,l}\Spaa{x_r,l}\Spbb{l,h_a}}{\Spaa{\xi,h_a}\Spaa{x_r,h_a}\Spaa{l,h_a}}.\Label{Eq:DoubleTraceMHV1-0}
\eea
On the second line, the explicit expression of $\epsilon_{h_a}\cdot k_l$ has been substituted and the expression in the square brackets is nothing but the contribution of the graph when the node $h_a$ is deleted.
Applying the above relation iteratively until all elements in the outer-graviton set $\mathsf{H}_B$ have been removed from the graph $\mathcal{F}$, we arrive
\bea
T^{\mathcal{F}}=T^{\mathcal{F}\setminus \mathsf{H}_B} \prod\limits_{{e(h_a,l)\in\,\mathcal{E}(\mathcal{G}')}}\frac{\Spaa{\xi,l}\Spaa{x_r,l}\Spbb{l,h_a}}{\Spaa{\xi,h_a}\Spaa{x_r,h_a}\Spaa{l,h_a}},\Label{Eq:DoubleTraceMHV1-1}
\eea
where $\mathcal{E}(\mathcal{G}')$ denotes the set of edges belonging to the spanning forest $\mathcal{G}'$ that is the collection of trees planted at nodes in $\left(\pmb{1}\setminus\{x_r\}\right)\cup\pmb{2}\cup\mathsf{H}_A$. These trees  in $\mathcal{G}'$ have the same structures with those in the corresponding refined graph $\mathcal{F}$, but the edges have the new meaning. After this step, the remaining refined graph $\mathcal{F}\setminus \mathsf{H}_B$ involves only the two traces and the inner gravitons (i.e. gravitons on the bridge between the two traces). Manipulations in this step is shown by \figref{Fig:DoubleTraceRecursion}.

{\bf Step-2:} Now we separate the two traces $\pmb{1}$ and $\pmb{2}$ (and the inner gravitons attached to each) from one another. Assume that the bridge between $\pmb{1}$ and $\pmb{2}$ in the remaining refined graph has the following structure
 \bea
 l_1=x_k\in \pmb{1}\setminus\{x_r\}\leftarrow c_1\leftarrow\dots\leftarrow c_{u-1}\leftarrow c(=c_u)\leftrightarrow d(=d_1)\rightarrow \dots\rightarrow d_v\rightarrow a\in\pmb{2}\setminus\{b\}, \Label{Eq:Bridge}
 \eea
where each type-2 line is expressed by $\leftarrow$ or $\rightarrow$, the type-3 line is denoted by $\leftrightarrow$, the gluon $l_1\in \pmb{1}$ is supposed to be $x_k$. According to the refined graphic rule, $T^{\mathcal{F}\setminus \mathsf{H}_B}$ in \eqref{Eq:DoubleTraceMHV1-1} is explicitly written as
\bea
T^{\mathcal{F}\setminus \mathsf{H}_B}&=&\Biggl[\,\Sl_{\shuffle}\frac{\Spaa{g_i,g_j}^4}{\left(x_1,\dots,x_k,\{x_{k+1},\dots,x_{r-1}\}\shuffle\{c_1,\dots,c=c_u,d=d_1,\dots,d_v,a,\pmb{\beta},b\},x_r\right)}\Biggr]\nn
&&~~~~~~~~~~~\times~\left(-k_c\cdot k_d\right)\,\prod\limits_{p=1}^{u}\epsilon_{c_p}\cdot k_{c_{p-1}}\,\prod\limits_{q=1}^{v}\left(-\epsilon_{d_q}\cdot k_{d_{q+1}}\right),\Label{Eq:DoubleTraceMHV1-2}
\eea
where $c_0\equiv x_k$ and $d_{v+1}\equiv a$. Note that the permutations $\{x_{k+1},\dots,x_{r-1}\}\shuffle\{c_1,\dots,c=c_u,d=d_1,\dots,d_v,a,\pmb{\beta},b\}$ can be reexpressed as
\bea
\bigl(\pmb{\gamma}\in\{x_{k+1},\dots,x_{r-1}\}\shuffle\{c_1,\dots,c=c_u\}\bigr)\shuffle\{ d=d_1,\dots,d_v,a,\pmb{\beta},b\}\big|_{c\prec d}.
\eea
 When the identity (\ref{Eq:PTRelation2}) is applied, the expression in the square brackets of \eqref{Eq:DoubleTraceMHV1-2} turns into
\bea
     &&\Sl_{\pmb{\gamma}} \frac{\Spaa{g_i,g_j}^4}{\left(x_1,\dots,x_k,\pmb{\gamma},x_r\right)}\frac{\Spaa{c,x_r}}{\langle c, d=d_1,\dots,d_v,a,\pmb{\beta},b,x_r\rangle}\Label{Eq:DoubleTraceMHV2}\nn
     &=&\Sl_{\pmb{\gamma}} \frac{\Spaa{g_i,g_j}^4}{\left(x_1,\dots,x_k,\pmb{\gamma},x_r\right)\left( d=d_1,\dots,d_v,a,\pmb{\beta},b\right)}\,\frac{\Spaa{b,d}\Spaa{c,x_r}}{\Spaa{c,d}\Spaa{b,x_r}}.
     \eea
Thus the contribution of two traces (and inner gravitons attached to them) in \eqref{Eq:DoubleTraceMHV1-2} are separated as follows
\bea
T^{\mathcal{F}\setminus \mathsf{H}_B}&=&\left(-k_c\cdot k_d\right)\frac{\Spaa{c,x_r}\Spaa{d,b}}{\Spaa{c,d}\Spaa{x_r,b}}\nn
&&\times\Biggl[\,\Sl_{\pmb{\gamma}}\frac{\Spaa{g_i,g_j}^4}{\left(x_1,\dots,x_k,\pmb{\gamma},x_r\right)}\prod\limits_{p=1}^{u}\epsilon_{c_p}\cdot k_{c_{p-1}}\Biggr]\Biggl[\,\frac{1}{\left( d=d_1,\dots,d_v,a,\pmb{\beta},b\right)}\prod\limits_{q=1}^{v}\left(-\epsilon_{d_q}\cdot k_{d_{q+1}}\right)\Biggr].\Label{Eq:DoubleTraceMHV3}
\eea
\begin{figure}
\centering
\includegraphics[width=1\textwidth]{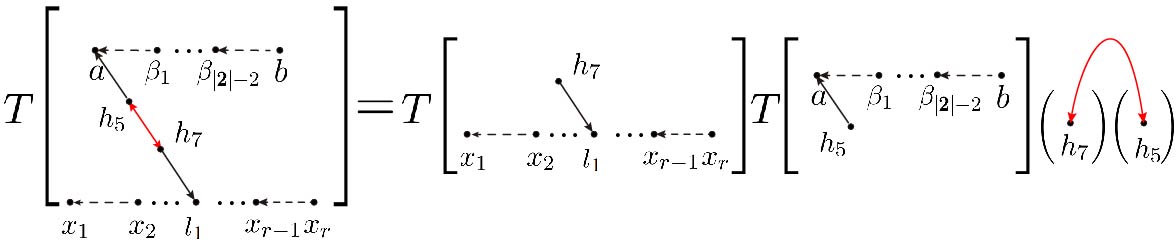}
\caption{An example for step-2 in the general proof of the  formula (\ref{Eq:DoubleTraceMHVgen1}). In  this step, the contribution of the remaining refined graph in \figref{Fig:DoubleTraceRecursion} is further decomposed such that the two traces and the inner gravitons attached to them are separated. After this step, the double-arrow line between $h_5$ and $h_7$ now stands for the crossing factor $\left(-k_{h_5}\cdot k_{h_7}\right)\frac{\Spaa{h_5,b}\Spaa{x_r,h_7}}{\Spaa{h_7,h_5}\Spaa{b,x_r}}$.  }\label{Fig:DoubleTraceRecursion2}
\end{figure}
Here, without loss of generality, the factor $\Spaa{g_i,g_j}^4$  is assumed to be absorbed into the Parke-Taylor factor involving the trace $\pmb{1}$. This step is demonstrated by \figref{Fig:DoubleTraceRecursion2}.

\begin{figure}
\centering
\includegraphics[width=1\textwidth]{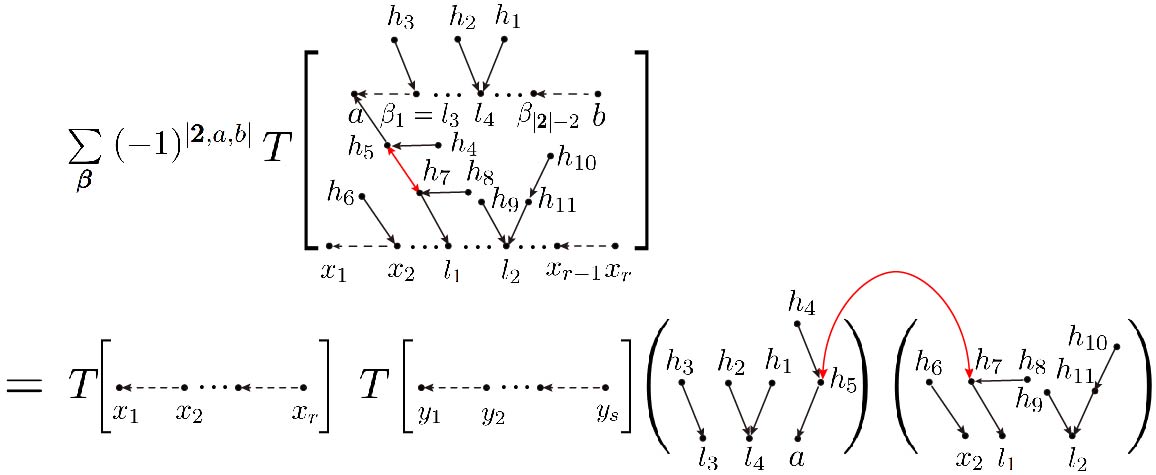}
\caption{An example for step-3 in the general proof of the  formula (\ref{Eq:DoubleTraceMHVgen1}). In this step, the inner gravitons $h_7$ and $h_5$ in \figref{Fig:DoubleTraceRecursion3} are further extracted from the traces $\pmb{1}$ and $\pmb{2}$ respectively. After summing over all $\pmb{\beta}\in \mathsf{KK}[\pmb{2},a,b]$ with the proper sign $(-1)^{|\pmb{2},a,b|}$, the second trace becomes a Parke-Taylor factor with the standard permutation $y_1,y_2,\dots,y_s$. Therefore, we get the structure (\ref{Eq:DoubleTraceMHV6}). Here the gluons in $\pmb{1}$ and $\pmb{2}$ are considered as roots, while gravitons $h_5$ and $h_7$ play as the inner gravitons. The bridge between the two traces is $l_1\leftarrow h_7\leftrightarrow h_5 \rightarrow a$. }\label{Fig:DoubleTraceRecursion3}
\end{figure}

{\bf Step-3:} We now extract the gravitons $c_1,\dots,c_u$ and $d_1,\dots,d_v$ from the traces $\pmb{1}$ and $\pmb{2}$ respectively.
The contribution of the trace $\pmb{1}$ (and the inner gravitons attached to it) is given by the expression inside the first pair of square brackets in \eqref{Eq:DoubleTraceMHV3}, which is further simplified as
\bea
&&\Sl_{\pmb{\gamma}}\frac{\Spaa{g_i,g_j}^4}{\left(x_1,\dots,x_k,\pmb{\gamma},x_r\right)}\prod\limits_{p=1}^{u}\epsilon_{c_p}\cdot k_{c_{p-1}}\nn
&=&\frac{\Spaa{g_i,g_j}^4}{\left(x_1,\dots,x_r\right)}\left(\prod\limits_{p=1}^u\frac{\Spaa{c_{p-1},x_r}}{\Spaa{c_{p-1},c_p}\Spaa{c_p,x_r}}\right)\left(\prod\limits_{p=1}^{u}\epsilon_{c_p}\cdot k_{c_{p-1}}\right)\nn
&=&\frac{\Spaa{g_i,g_j}^4}{\left(x_1,\dots,x_r\right)}\prod\limits_{p=1}^{u}\frac{\Spaa{\xi,c_{p-1}}\Spaa{x_r,c_{p-1}}\Spbb{c_{p-1},c_p}}{\Spaa{\xi,c_{p}}\Spaa{x_r,c_p}\Spaa{c_{p-1},c_p}},\Label{Eq:DoubleTraceMHV4}
\eea
where permutations $\pmb{\gamma}$ satisfy $\pmb{\gamma}\in\{x_{k+1},\dots,x_{r-1}\}\shuffle\{c_1,\dots,c=c_u\}$. For a given $c_p$ in \eqref{Eq:DoubleTraceMHV4}, the identity (\ref{Eq:PTRelation1}) has been applied and the explicit expression of $\epsilon_{c_p}\cdot k_{c_{p-1}}$ has been inserted.
For the trace $\pmb{2}$ in \eqref{Eq:DoubleTraceMHV3}, we extract gravitons $d_1,\dots, d_v$ out from the corresponding Parke-Taylor factor (noting that the Parke-Taylor factor has cyclic symmetry) as follows
\bea
&&\frac{1}{\left( d=d_1,\dots,d_v,a,\pmb{\beta},b\right)}\prod\limits_{q=1}^{v}\left(-\epsilon_{d_q}\cdot k_{d_{q+1}}\right)\nn
&=& \frac{1}{\left(a,\pmb{\beta},b\right)} \left(\prod\limits_{q=1}^v\frac{\Spaa{b,d_{q+1}}}{\Spaa{b,d_{q}}\Spaa{d_{q},d_{q+1}}}\right)\prod\limits_{q=1}^{v}\left(-\epsilon_{d_q}\cdot k_{d_{q+1}}\right)\nn
&=&\frac{1}{\left(a,\pmb{\beta},b\right)}\prod\limits_{q=1}^{v}\frac{\Spaa{\xi,d_{q+1}}\Spaa{b,d_{q+1}}\Spbb{d_{q+1},d_q}}{\Spaa{\xi,d_q}\Spaa{b,d_{q}}\Spaa{d_{q+1},d_{q}}}.\Label{Eq:DoubleTraceMHV5}
\eea
When we sum over all possible permutations $\pmb{\beta}\in \mathsf{KK}[\pmb{2},a,b]$, each of which is dressed by the sign $(-1)^{|\pmb{2},a,b|}$, the Parke-Taylor factors ${1}\setminus{\left(a,\pmb{\beta},b\right)}$ turn into the one with the standard permutation $y_1,\dots,y_s$, because of the KK relation (\ref{Eq:KK}).  We thus get
\bea
\frac{1}{\left(y_1,\dots,y_s\right)}\prod\limits_{q=1}^{v}\frac{\Spaa{\xi,d_{q+1}}\Spaa{b,d_{q+1}}\Spbb{d_{q+1},d_q}}{\Spaa{\xi,d_q}\Spaa{b,d_{q}}\Spaa{d_{q+1},d_{q}}}.\Label{Eq:DoubleTraceMHV5-1}
\eea
Eq. (\ref{Eq:DoubleTraceMHV3}), \eqref{Eq:DoubleTraceMHV4} and \eqref{Eq:DoubleTraceMHV5-1} together imply that  \eqref{Eq:DoubleTraceMHV1-1} can be reexpressed as (shown by \figref{Fig:DoubleTraceRecursion3})
\bea
&&\frac{\Spaa{g_i,g_j}^4}{\left(x_1,\dots,x_r\right)\,\left(y_1,\dots,y_s\right)}\left(-k_c\cdot k_d\right)\frac{\Spaa{c,x_r}\Spaa{d,b}}{\Spaa{c,d}\Spaa{x_r,b}}\,\prod\limits_{p=1}^{u}\,\frac{\Spaa{\xi,c_{p-1}}\Spaa{x_r,c_{p-1}}\Spbb{c_{p-1},c_p}}{\Spaa{\xi,c_{p}}\Spaa{x_r,c_p}\Spaa{c_{p-1},c_p}}\nn
&&~~~~~~~~~~~~~~~~~~~~\times\,\prod\limits_{q=1}^{v}\,\frac{\Spaa{\xi,d_{q+1}}\Spaa{b,d_{q+1}}\Spbb{d_{q+1},d_q}}{\Spaa{\xi,d_q}\Spaa{b,d_{q}}\Spaa{d_{q+1},d_{q}}}\prod\limits_{{e(h_a,l)\in\,\mathcal{E}(\mathcal{G}')}}\,\frac{\Spaa{\xi,l}\Spaa{x_r,l}\Spbb{l,h_a}}{\Spaa{\xi,h_a}\Spaa{x_r,h_a}\Spaa{l,h_a}}\Label{Eq:DoubleTraceMHV6}
\eea
when all possible $\pmb{\beta}\in \mathsf{KK}[\pmb{2},a,b]$ are summed over.
This is exactly one term of \eqref{Eq:DoubleTraceMHVgen2}, while the inner graviton set is  $\mathsf{H}_A=\{c_1,\cdots,c_u,d_1,\cdots,d_v\}$, the bridge $\mathcal{B}(H_A)$ between the two traces is given by (\ref{Eq:Bridge}) and the choice of gauge  is
\bea
 \chi=b\,,~~\zeta=\eta=x_r.\Label{Eq:GaugeGenProof}
\eea
After summing over all splits $\mathsf{H}\to \mathsf{H}_A,\mathsf{H}_B$, all possible bridges (\ref{Eq:Bridge}) for a given split $\mathsf{H}\to \mathsf{H}_A,\mathsf{H}_B$ and all spanning forests rooted at $\left(\pmb{1}\setminus\{x_r\}\right)\cup\pmb{2}\cup\mathsf{H}_A$ for a given bridge, we finally get \eqref{Eq:DoubleTraceMHVgen2} (thus \eqref{Eq:DoubleTraceMHVgen1}) with the choice of gauge (\ref{Eq:GaugeGenProof}) (noting  that the terms in \eqref{Eq:DoubleTraceMHVgen2} which invlove trees planted on $x_r$ and/or the bridge (\ref{Eq:Bridge}) with $a=b$  must vanish under this choice of gauge, due to the antisymmetry of spinor products). Since we have already proven that \eqref{Eq:DoubleTraceMHVgen2} is independent of the choice of gauge, the general proof of  \eqref{Eq:DoubleTraceMHVgen2} (thus \eqref{Eq:DoubleTraceMHVgen1}) has been completed.

\section{The vanishing configurations}\label{sec:VanishingAmplitude}

In this section, we investigate the vanishing amplitudes with two negative-helicity particles. We first introduce several properties of refined graphs {\emph{when the corresponding YM amplitudes in the expansion (\ref{Eq:PureYMExpansion}) are MHV amplitudes (which are expressed by Parke-Taylor formula).}} Using these properties, we prove that the double-trace amplitudes with the $(h_i^-,g_j^-)$ configuration, the single- and double- trace amplitudes with the $(h_i^-,h_j^-)$ configuration as well as all amplitudes with arbitrary two negative-helicity particles and more than two traces have to vanish.

\subsection{Helpful properties of refined graphs for amplitudes with two negative-helicity particles}
Now we present several usful properties of the refined graphs when the amplitude has two negative-helicity particles. \emph{Our choice of gauge is always the standard one: i.e. all the reference momenta of positive-helicity gravitons are chosen as $k^{\mu}_{h_i}$ ($h_i$ is a negative-helicity graviton), while the reference momenta of all negative-helicity gravitons, i.e. the $h_i$ for the $(h_i^-,g_j^-)$ configuration and $h_i$, $h_j$ for the $(h_i^-,h_j^-)$ configuration, are chosen as $\xi^{\mu}$.} With this choice of gauge, the refined graphs satisfy the following properties.

\begin{figure}
\centering
\includegraphics[width=0.9\textwidth]{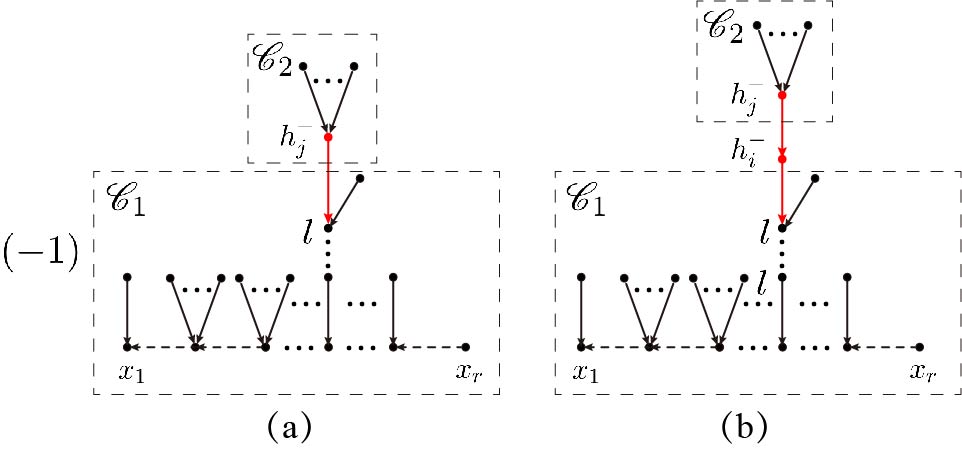}
\caption{Graphs with the structures (a) and (b) cancel with each other if the corresponding amplitudes are described by Parke-Taylor factors.
}\label{Fig:Cancellation1}
\end{figure}

{\bf {Property-1:}} Suppose $h_i$ is a negative-helicity graviton  and $\{\mathcal{F}'\}$ is the set of graphs satisfying the following two conditions: (i). The $h_i$ is a leaf which starts a type-2 line $\epsilon_{h_i}\cdot k_l$, (ii). All graphs in $\{\mathcal{F}'\}$ are reduced to an identical graph $\mathcal{F}$ when $h_i$ is deleted. {\bf\emph{The sum of contributions of all graphs in $\{\mathcal{F}'\}$ is reduced to the contribution of $\mathcal{F}=\mathcal{F}'\setminus\{h_a\}$ with an extra minus.}} This property, when the graphs $\mathcal{F}'$ involve only one component (see \figref{Fig:SingleTraceConfiguration2}), has already been proven in \secref{sec:SingleTraceConf2}. Proof for the cases with more general structure follows from a similar statement.

{\bf {Property-2:}} {\bf\emph{Graphs of the patterns \figref{Fig:Cancellation1} (a) and (b) must cancel with each other.}} This is shown as follows.  When we extract the gravitons from the Parke-Taylor factors, the contributions of $\mathscr{C}_1$ ($\mathscr{C}_2$) are the same for distinct graphs \figref{Fig:Cancellation1} (a) and (b). Thus we only need to compare the contribution of the substructures colored red. In \figref{Fig:Cancellation1} (a), the red part contributes a factor
\bea
(-1)(\epsilon_{h_j}^-\cdot k_l)\,\left(\frac{\Spaa{l,x_r}}{\Spaa{l,h_j}\Spaa{h_j,x_r}}\right),
\eea
where the first factor comes from the definition of the type-2 line and the second factor is obtained when extracting $h_j$ out from the Parke-Taylor factor, $(-1)$ is the overall sign in \figref{Fig:Cancellation1} (a). When the first factor is expressed by spinor products, the above expression is reduced into
\bea
-\frac{\Spaa{l,h_j}\Spbb{\xi,l}}{\Spbb{\xi,h_j}}\frac{\Spaa{l,x_r}}{\Spaa{l,h_j}\Spaa{h_j,x_r}}=-\frac{\Spaa{l,x_r}\Spbb{\xi,l}}{\Spaa{h_j,x_r}\Spbb{\xi,h_j}}.
\eea
The red part of  \figref{Fig:Cancellation1} (b) contributes the following factor
\bea
&&(\epsilon_{h_j}^-\cdot k_{h_i}) \frac{\Spaa{h_i,x_r}}{\Spaa{h_i,h_j}\Spaa{h_j,x_r}} (\epsilon_{h_i}^-\cdot k_l)\frac{\Spaa{l,x_r}}{\Spaa{l,h_i}\Spaa{h_i,x_r}}\nn
&=&\frac{\Spaa{h_i,h_j}\Spbb{\xi,h_i}}{\Spbb{\xi,h_j}}\frac{\Spaa{h_i,x_r}}{\Spaa{h_i,h_j}\Spaa{h_j,x_r}}\frac{\Spaa{l,h_i}\Spbb{\xi,l}}{\Spbb{\xi,h_i}}\frac{\Spaa{l,x_r}}{\Spaa{l,h_i}\Spaa{h_i,x_r}}\nn
&=&\frac{\Spaa{l,x_r}\Spbb{\xi,l}}{\Spaa{h_j,x_r}\Spbb{\xi,h_j}},
\eea
which is just the corresponding factor in \figref{Fig:Cancellation1} (a) with an opposite sign. Thus \figref{Fig:Cancellation1} (a) and (b)  cancel with each other.

\begin{figure}
\centering
\includegraphics[width=0.98\textwidth]{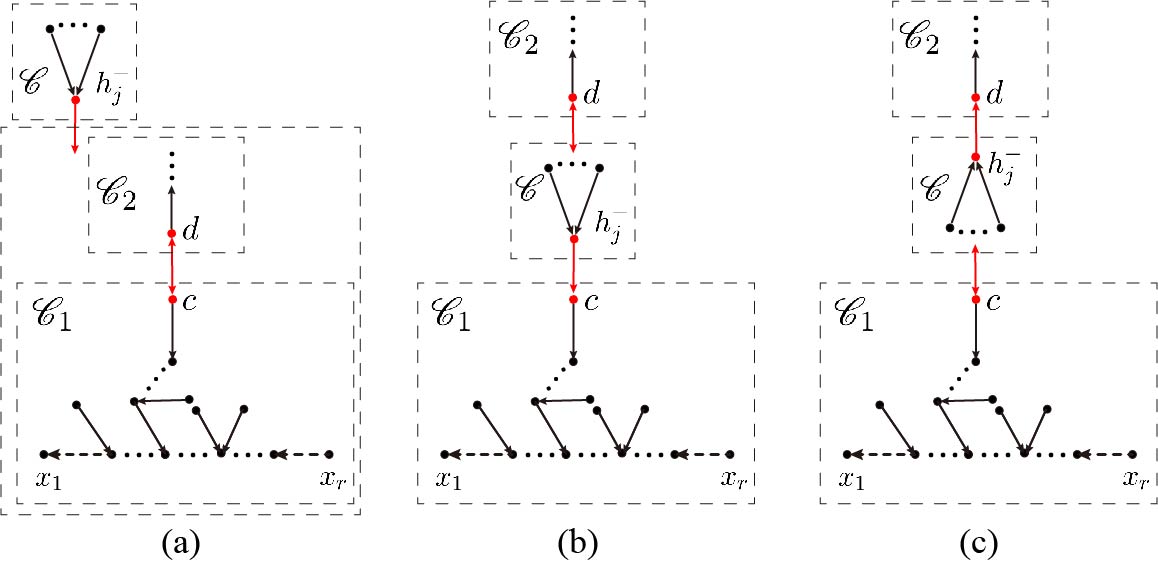}
\caption{Contributions of graphs with the typical structures (a), (b) and (c) cancel with each other.}\label{Fig:Cancellation3}
\end{figure}

{\bf {Property-3:}} let $\mathscr{C}$ be a tree structure where all lines are type-2 lines whose arrows point towards the negative-helicity graviton $h_j$, $\mathscr{C}_1$ and $\mathscr{C}_2$ be other two tree structures. Suppose the trace $\pmb{1}=\{x_1,\dots,x_r\}$ belongs to $\mathscr{C}_1$. {\bf\emph{Graphs with the structures \figref{Fig:Cancellation3} (a), (b) and (c) must cancel with each other.}} In \figref{Fig:Cancellation3}, a line connected to a boxed region but not a concrete node always means we sum over graphs where the line is connected to all nodes  (except $x_r\in \pmb{1}$) in that region. To prove the cancellation between these three graphs, we only need to consider the $\mathscr{C}$ part and the red lines in each graph because  either the $\mathscr{C}_1$ or the $\mathscr{C}_2$  part provides an identical expression in all three graphs.
\begin{itemize}
\item {\bf(a).} In \figref{Fig:Cancellation3} (a), when extracting the $\mathscr{C}$ part from the Parke-Taylor factors
we get a factor of the following form
  \bea
 \left[(\epsilon_{h_j}\cdot k_l)\frac{\Spaa{l,x_r}}{\Spaa{l,h_j}\Spaa{h_j,x_r}}\right]\left[\prod\limits_{\substack{e(x,y)\in\mathcal{E}(\mathscr{C})}}(\epsilon_{x}\cdot k_y)\frac{\Spaa{y,x_r}}{\Spaa{y,x}\Spaa{x,x_r}}\right],
  \eea
where $l\in(\mathscr{C}_1\setminus\{x_r\})\cup\mathscr{C}_2$ is supposed to be the end node of the type-2 line which starts at $h_j$. We extract the node $d$ after extracting all other nodes in $\mathscr{C}_2$ part from the Parke-Taylor factors. Then the node $d$ is associated with the following factor
\bea
(-k_d\cdot k_c)\frac{\Spaa{c,x_r}}{\Spaa{c,d}\Spaa{d,x_r}}.
\eea
Putting the above factors together and summing over $l\in(\mathscr{C}_1\setminus\{x_r\})\cup\mathscr{C}_2$ we get
\bea
T_1&=&\Sl_{l\in(\mathscr{C}_1\setminus\{x_r\})\cup\mathscr{C}_2}\left[(\epsilon_{h_j}\cdot k_l)\frac{\Spaa{l,x_r}}{\Spaa{l,h_j}\Spaa{h_j,x_r}}\right]\left[\prod\limits_{\substack{e(x,y)\in\mathcal{E}(\mathscr{C})}}(\epsilon_{x}\cdot k_y)\frac{\Spaa{y,x_r}}{\Spaa{y,x}\Spaa{x,x_r}}\right](-k_d\cdot k_c)\frac{\Spaa{c,x_r}}{\Spaa{c,d}\Spaa{d,x_r}}\nn
&=&\Sl_{l\in\mathscr{C}_1\cup\mathscr{C}_2}\frac{1}{2}\frac{\Spbb{\xi,l}\Spaa{l,x_r}}{\Spbb{\xi,h_j}\Spaa{h_j,x_r}}\left[\prod\limits_{\substack{e(x,y)\in\mathcal{E}(\mathscr{C})}}(\epsilon_{x}\cdot k_y)\frac{\Spaa{y,x_r}}{\Spaa{y,x}\Spaa{x,x_r}}\right]\frac{\Spbb{c,d}\Spaa{c,x_r}}{\Spaa{d,x_r}}\nn
&=&-\Sl_{l\in\mathscr{C}}\frac{1}{2}\frac{\Spbb{\xi,l}\Spbb{c,d}}{\Spbb{\xi,h_j}}\left[\frac{\Spaa{l,x_r}\Spaa{c,x_r}}{\Spaa{h_j,x_r}\Spaa{d,x_r}}\right]\left[\prod\limits_{\substack{e(x,y)\in\mathcal{E}(\mathscr{C})}}(\epsilon_{x}\cdot k_y)\frac{\Spaa{y,x_r}}{\Spaa{y,x}\Spaa{x,x_r}}\right],
\eea
in which, $e(x,y)\in\mathcal{E}(\mathscr{C})$ denoted the set of (type-2) lines in $\mathscr{C}$ and $e(x,y)$ denoted the type-2 line starting at $x$ and ending at $y$,
the term with $l=x_r$ was included in the second line because of $\Spaa{x_r,x_r}=0$, momentum conservation was applied on the third line.

\item {\bf(b).} In \figref{Fig:Cancellation3} (b), when nodes in the $\mathscr{C}$ part and the node $d$ are extracted out from the Parke-Taylor factors, we get the following factor
\bea
T_2&=&\left[\Sl_{l\in\mathscr{C}}(-k_d\cdot k_l)\frac{\Spaa{l,x_r}}{\Spaa{l,d}\Spaa{d,x_r}}\right]\left[(\epsilon_{h_j}\cdot k_c)\frac{\Spaa{c,x_r}}{\Spaa{c,h_j}\Spaa{h_j,x_r}}\right]\left[\prod\limits_{\substack{e(x,y)\in\mathcal{E}(\mathscr{C})}}(\epsilon_{x}\cdot k_y)\frac{\Spaa{y,x_r}}{\Spaa{y,x}\Spaa{x,x_r}}\right]\nn
&=&\Sl_{l\in\mathscr{C}}\frac{1}{2}\frac{\Spbb{l,d}\Spbb{\xi,c}}{\Spbb{\xi,h_j}}\left[\frac{\Spaa{l,x_r}\Spaa{c,x_r}}{\Spaa{d,x_r}\Spaa{h_j,x_r}}\right]\left[\prod\limits_{\substack{e(x,y)\in\mathcal{E}(\mathscr{C})}}(\epsilon_{x}\cdot k_y)\frac{\Spaa{y,x_r}}{\Spaa{y,x}\Spaa{x,x_r}}\right],
\eea
where all possible $l\in\mathscr{C}$ were summed over.

\item {\bf(c).} To evaluate  \figref{Fig:Cancellation3} (c), we suppose the node $l\in\mathscr{C}$ is connected to $c$ via the type-3 line. Assuming that nodes on the path from $l$ to $h_j$ are $u_1$, $u_2$, ..., $u_k$, respectively, all the type-2 lines between adjacent nodes on this path point away from the trace $\pmb{1}$. When extracting the nodes $u_1$, ..., $u_k$, $h_j$, from the Parke-Taylor factors according to \eqref{Eq:PTRelation2}, we get a factor
    \bea
    &&(-\epsilon_{u_k}\cdot k_{h_j})(-\epsilon_{u_{k-1}}\cdot k_{u_{k}})\dots (-\epsilon_{u_1}\cdot k_{u_2})(-\epsilon_{l}\cdot k_{u_1})\frac{\Spaa{l,x_r}}{\Spaa{l,u_1,\dots,u_k,h_j,x_r}}\nn
    &=&(\epsilon_{u_k}\cdot k_{h_j})\,(\epsilon_{u_{k-1}}\cdot k_{u_{k}})\,\dots \,(\epsilon_{u_1}\cdot k_{u_2})\,(\epsilon_{l}\cdot k_{u_1})\,\frac{\Spaa{h_j,x_r}}{\Spaa{h_j,u_k,\dots,u_1,l,x_r}}\left(\frac{\Spaa{l,x_r}}{\Spaa{h_j,x_r}}\right)^2\nn
    &=&\biggl[(\epsilon_{u_k}\cdot k_{h_j})\frac{\Spaa{h_j,x_r}}{\Spaa{h_j,u_k}\Spaa{u_k,x_r}}(\epsilon_{u_{k-1}}\cdot k_{u_{k}})\frac{\Spaa{u_k,x_r}}{\Spaa{u_k,u_{k-1}}\Spaa{u_{k-1},x_r}}\nn
    &&~~~~~~~~~~~~~~~~\dots(\epsilon_{u_1}\cdot k_{u_2})\frac{\Spaa{u_2,x_r}}{\Spaa{u_2,u_1}\Spaa{u_1,x_r}}(\epsilon_{l}\cdot k_{u_1})\frac{\Spaa{u_1,x_r}}{\Spaa{u_1,l}\Spaa{l,x_r}}\biggr]\left(\frac{\Spaa{l,x_r}}{\Spaa{h_j,x_r}}\right)^2,\nonumber
    \eea
where the antisymmetry of spinor products has been applied. The expression in the square brackets on the last line would have the form as if arrows on this path pointed towards the trace $\pmb{1}$. Arrows of lines in $\mathscr{C}$ which do not on the path from $l$ to $h_j$ already point towards the trace $\pmb{1}$. Hence full contribution of lines in the $\mathscr{C}$ part reads
\bea
\left[\prod\limits_{\substack{e(x,y)\in\mathcal{E}(\mathscr{C})}}(\epsilon_{x}\cdot k_y)\frac{\Spaa{y,x_r}}{\Spaa{y,x}\Spaa{x,x_r}}\right]\left(\frac{\Spaa{l,x_r}}{\Spaa{h_j,x_r}}\right)^2.
\eea
The node $d$ and the node $l$ respectively provide the following factors when they are extracted out
\bea
&&(-\epsilon_{h_j}\cdot k_{d})\frac{\Spaa{h_j,x_r}}{\Spaa{h_j,d}\Spaa{d,x_r}},~~~~~~~(-k_c\cdot k_l)\frac{\Spaa{c,x_r}}{\Spaa{c,l}\Spaa{l,x_r}}.
\eea
\begin{figure}
\centering
\includegraphics[width=0.98\textwidth]{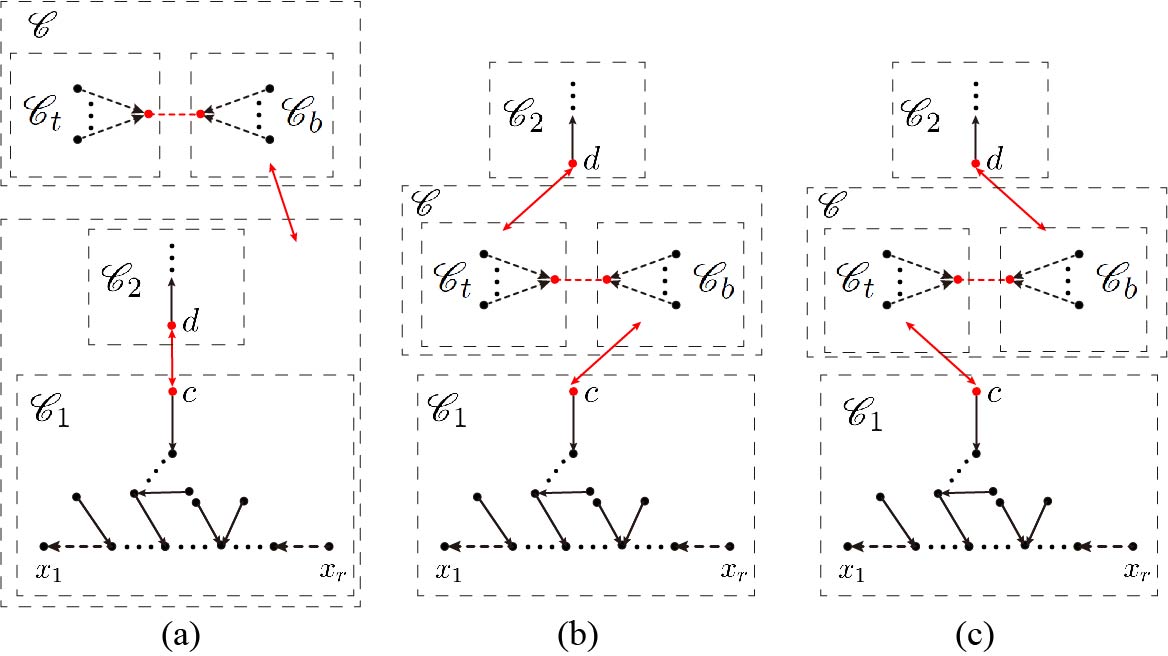}
\caption{Contributions of graphs with the above typical structures cancel out. Here, the substructure $\mathscr{C}$ is divided into two parts $\mathscr{C}_t$ and $\mathscr{C}_b$ by a dashed line with no arrow.}\label{Fig:Cancellation4}
\end{figure}
Therefore, the contributions of lines in $\mathscr{C}$ together with the red parts in \figref{Fig:Cancellation3} (c) are collected as
\bea
(-\epsilon_{h_j}\cdot k_{d})\frac{\Spaa{h_j,x_r}}{\Spaa{h_j,d}\Spaa{d,x_r}}(-k_c\cdot k_l)\frac{\Spaa{c,x_r}}{\Spaa{c,l}\Spaa{l,x_r}}\left[\prod\limits_{\substack{e(x,y)\in\mathcal{E}(\mathscr{C})}}(\epsilon_{x}\cdot k_y)\frac{\Spaa{y,x_r}}{\Spaa{y,x}\Spaa{x,x_r}}\right]\left(\frac{\Spaa{l,x_r}}{\Spaa{h_j,x_r}}\right)^2.
\eea
When spinor expressions are applied and $l\in\mathscr{C}$ is summed over, the above expression turns into
\bea
T_3=\Sl_{l\in\mathscr{C}}\frac{1}{2}\frac{\Spbb{c,l}\Spbb{\xi,d}}{\Spbb{\xi,h_j}}\left[\frac{\Spaa{l,x_r}\Spaa{c,x_r}}{\Spaa{d,x_r}\Spaa{h_j,x_r}}\right]\left[\prod\limits_{\substack{e(x,y)\in\mathcal{E}(\mathscr{C})}}(\epsilon_{x}\cdot k_y)\frac{\Spaa{y,x_r}}{\Spaa{y,x}\Spaa{x,x_r}}\right].
\eea
\end{itemize}
The sum of $T_1$, $T_2$ and $T_3$ for any $l$ precisely cancel out due to Schouten identity (\ref{Eq:SpinorProp2}).

{\bf{Property-4:}} {\bf\emph{Graphs  with the structures \figref{Fig:Cancellation4} (a), (b) and (c) cancel with each other.}} Before proving this property, let us first explain the meaning of dashed lines in the graphs: \emph{(1)} the dashed arrow lines inherit the definition of the type-4 line (see \figref{Fig:LineStyles} (d)) but the two ends of these lines can be  gravitons and/or gluons, \emph{(2).} the dashed line with no arrow in \figref{Fig:Cancellation4} (colored red) also reflects the relative position between its two end nodes, but it does not bring any sign. The cancellation between graphs \figref{Fig:Cancellation4} (a), (b) and (c) only relies on the relative positions between nodes and possible signs caused by the arrows (just as the property-3). Let us now evaluate the different parts in the three graphs in turn.
\begin{figure}
\centering
\includegraphics[width=0.85\textwidth]{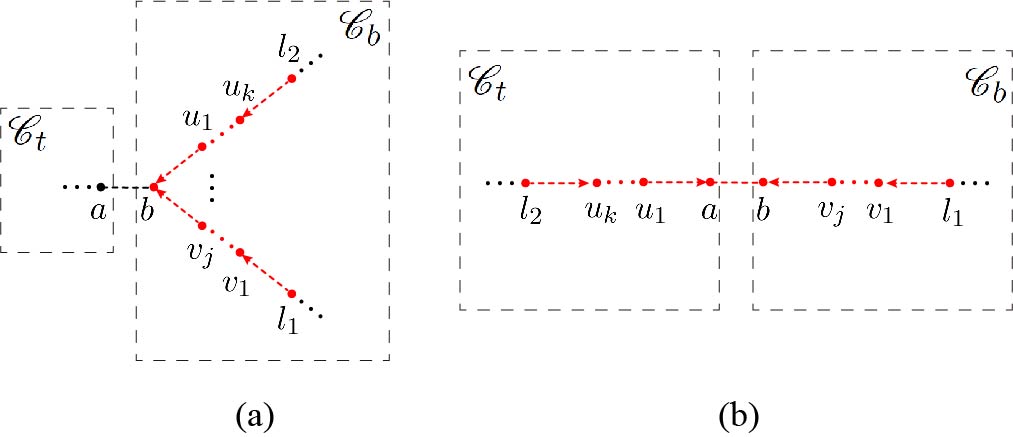}
\caption{(a): the path between $l_1$ and $l_2$ in \eqref{Eq:Cancellation1}, (b): the path between $l_2$ and $l_1$ in \figref{Fig:Cancellation5_2} (c)}\label{Fig:Cancellation5_2}
\end{figure}
\begin{itemize}
\item {\bf(a)} Suppose that the two end nodes of the type-3 line between $\mathscr{C}_b$ and $\mathscr{C}_1\cup\mathscr{C}_2$ in \figref{Fig:Cancellation4} (a) are respectively $l_1\in \mathscr{C}_b$ and $l_2\in(\mathscr{C}_1\setminus\{x_r\})\cup\mathscr{C}_2$. Then all lines in $\mathscr{C}$, the type-3 line between $l_1$ , $l_2$ and the type-3 line between $c$, $d$ together produce the following factor
\bea
\Sl_{l_2\in(\mathscr{C}_1\setminus\{x_r\})\cup\mathscr{C}_2}\Sl_{l_1\in\mathscr{C}_b}T[\mathscr{C}(l_1)] \left[(-k_{l_1}\cdot k_{l_2})\frac{\Spaa{l_2,x_r}}{\Spaa{l_2,l_1}\Spaa{l_1,x_r}}\right]\left[(-k_d\cdot k_c)\frac{\Spaa{c,x_r}}{\Spaa{c,d}\Spaa{d,x_r}}\right],\Label{Eq:Cancellation0}
\eea
where $T[\mathscr{C}(l_1)]$ denotes the full contributions of lines in the region $\mathscr{C}$ when $l_1$ is the node nearest to the trace $\pmb{1}$. Momentum conservation allows one to convert the summation over $l_2\in(\mathscr{C}_1\setminus\{x_r\})\cup\mathscr{C}_2$ into a summation over $l_2\in\mathscr{C}_t\cup\mathscr{C}_b$ with an extra sign $(-1)$, which further splits into the summations over $l_2\in\mathscr{C}_b$ and $l_2\in\mathscr{C}_t$. The sum of terms with  $l_2\in\mathscr{C}_b$ is explicitly written as
\bea
-\Sl_{l_1,l_2\in\mathscr{C}_b}T[\mathscr{C}(l_1)] \left[(-k_{l_1}\cdot k_{l_2})\frac{\Spaa{l_2,x_r}}{\Spaa{l_2,l_1}\Spaa{l_1,x_r}}\right]\left[(-k_d\cdot k_c)\frac{\Spaa{c,x_r}}{\Spaa{c,d}\Spaa{d,x_r}}\right].\Label{Eq:Cancellation1}
\eea
For a given $l_1,l_2\in\mathscr{C}_b$ in the above expression, one can always find out another term in the summation by exchanging the roles of $l_1$ and $l_2$. The only difference between $T[\mathscr{C}(l_1)]$ and $T[\mathscr{C}(l_2)]$  is the factor corresponding to the path between $l_1$ and $l_2$ (see \figref{Fig:Cancellation5_2} (a)). Supposing the nodes between $l_1$ and $l_2$ are in turn $l_1,v_1,\dots,v_j, b, u_1,\dots,u_k,l_2$, the lines between $l_1$ and $l_2$ contribute the following factor to $T[\mathscr{C}(l_1)]$:
\bea
(-1)^{j+1}\frac{\Spaa{l_1,x_r}}{\Spaa{l_1,v_1,\dots,v_j,b,u_1,\dots,u_k,l_2,x_r}}=\frac{\Spaa{l_1,x_r}}{\Spaa{b,v_j,\dots,v_1,l_1}\Spaa{b,u_1,\dots,u_k,l_2,x_r}},\Label{Eq:Cancellation2}
\eea
when \eqref{Eq:PTRelation2} is been applied. Together with the second factor in \eqref{Eq:Cancellation1}, the above factor provides
\bea
&&\frac{\Spaa{l_1,x_r}}{\Spaa{b,v_j,\dots,v_1,l_1}\Spaa{b,u_1,\dots,u_k,l_2,x_r}}(-k_{l_1}\cdot k_{l_2})\frac{\Spaa{l_2,x_r}}{\Spaa{l_2,l_1}\Spaa{l_1,x_r}}\nn
&=&\frac{1}{2}\frac{\Spbb{l_2,l_1}}{\Spaa{b,v_j,\dots,v_1,l_1}\Spaa{b,u_1,\dots,u_k,l_2}}.
\eea
While exchanging the roles of $l_1$ and $l_2$, we get the same expression with an opposite sign. This antisymmetry indicates that \eqref{Eq:Cancellation1} vanishes when all $l_1,l_2\in \mathscr{C}_b$ are summed over. The expression (\ref{Eq:Cancellation0}) is then reduced into
\bea
T_1&=&-\Sl_{l_2\in\mathscr{C}_t}\Sl_{l_1\in\mathscr{C}_b}T[\mathscr{C}(l_1)] \left[(-k_{l_1}\cdot k_{l_2})\frac{\Spaa{l_2,x_r}}{\Spaa{l_2,l_1}\Spaa{l_1,x_r}}\right]\left[(-k_d\cdot k_c)\frac{\Spaa{c,x_r}}{\Spaa{c,d}\Spaa{d,x_r}}\right]\Label{Eq:Cancellation2}\nn
&=&\Sl_{l_2\in\mathscr{C}_t}\Sl_{l_1\in\mathscr{C}_b}\frac{1}{2}\Spbb{l_1,l_2}\Spbb{c,d}\frac{\Spaa{l_2,x_r}\Spaa{c,x_r}}{\Spaa{l_1,x_r}\Spaa{d,x_r}}T[\mathscr{C}(l_1)].
\eea

\item {\bf(b)} In \figref{Fig:Cancellation4} (b), the two type-3 lines and the $\mathscr{C}$ part together contribute a factor
\bea
T_2&=&\Sl_{l_2\in\mathscr{C}_t}\Sl_{l_1\in\mathscr{C}_b}(-k_d\cdot k_{l_2})\frac{\Spaa{l_2,x_r}}{\Spaa{l_2,d}\Spaa{d,x_r}}(-k_c\cdot k_{l_1})\frac{\Spaa{c,x_r}}{\Spaa{c,l_1}\Spaa{l_1,x_r}}T[\mathscr{C}(l_1)]\nn
&=&\Sl_{l_2\in\mathscr{C}_t}\Sl_{l_1\in\mathscr{C}_b}\frac{1}{2}\Spbb{l_2,d}\Spbb{c,l_1}\frac{\Spaa{l_2,x_r}\Spaa{c,x_r}}{\Spaa{l_1,x_r}\Spaa{d,x_r}}T[\mathscr{C}(l_1)].
\eea
\item {\bf(c)} In \figref{Fig:Cancellation4} (c), the two type-3 lines and the $\mathscr{C}$ part contribute the following factor
\bea
T_3&=&\Sl_{l_2\in\mathscr{C}_t}\Sl_{l_1\in\mathscr{C}_b}(-k_d\cdot k_{l_1})\frac{\Spaa{l_1,x_r}}{\Spaa{l_1,d}\Spaa{d,x_r}}(-k_c\cdot k_{l_2})\frac{\Spaa{c,x_r}}{\Spaa{c,l_2}\Spaa{l_2,x_r}} T[\mathscr{C}(l_2)]\nn
&=&\Sl_{l_2\in\mathscr{C}_t}\Sl_{l_1\in\mathscr{C}_b}\frac{1}{2}\Spbb{l_1,d}\Spbb{c,l_2}\frac{\Spaa{l_1,x_r}\Spaa{c,x_r}}{\Spaa{l_2,x_r}\Spaa{d,x_r}}T[\mathscr{C}(l_2)].
\eea
To relate the $T[\mathscr{C}(l_2)]$, which is the expression of $\mathscr{C}$ part while $l_2$ is the node nearest to the trace $\pmb{1}$, to $T[\mathscr{C}(l_1)]$, we note that the only difference between $T[\mathscr{C}(l_1)]$  and $T[\mathscr{C}(l_2)]$ is the path between the nodes $l_1\in\mathscr{C}_b$ and $l_2\in\mathscr{C}_t$ (see \figref{Fig:Cancellation5_2} (b)). In $T[\mathscr{C}(l_2)]$, this path  has the form
\bea
(-1)^{k+1}\frac{\Spaa{l_2,x_r}}{\Spaa{l_2,u_k,\dots,u_1,a,b,v_1,\dots,v_j,l_1,x_r}}=\frac{\Spaa{l_2,x_r}}{\Spaa{a,u_1,\dots,u_k,l_2}\Spaa{a,b}\Spaa{b,v_1,\dots,v_j,l_1,x_r}},\Label{Eq:Cancellation3}
\eea
where nodes on this path were  supposed to be $l_2,u_k,\dots,u_1,a,b,v_1,\dots,v_j,l_1$ in turn. Similarly, in $T[\mathscr{C}(l_1)]$, the path from $l_1$ to $l_2$  contributes a factor
\bea
\frac{\Spaa{l_1,x_r}}{\Spaa{b,v_1,\dots,v_j,l_1}\Spaa{b,a}\Spaa{a,u_1,\dots,u_k,l_2,x_r}}.\Label{Eq:Cancellation4}
\eea
Therefore, we arrive $T[\mathscr{C}(l_2)]=\left(-\frac{\Spaa{l_2,x_r}^2}{\Spaa{l_1,x_r}^2}\right)T[\mathscr{C}(l_1)]$ and $T_3$ is immediately rewritten as
\bea
T_3=\Sl_{l_2\in\mathscr{C}_t}\Sl_{l_1\in\mathscr{C}_b}\frac{1}{2}\Spbb{d,l_1}\Spbb{c,l_2}\frac{\Spaa{l_2,x_r}\Spaa{c,x_r}}{\Spaa{l_1,x_r}\Spaa{d,x_r}}T[\mathscr{C}(l_1)].
\eea

\end{itemize}
When summing $T_1$, $T_2$ and $T_3$ together, we get
\bea
T_1+T_2+T_3=\Sl_{l_2\in\mathscr{C}_t}\Sl_{l_1\in\mathscr{C}_b}\frac{1}{2}\left(\Spbb{l_1,l_2}\Spbb{c,d}+\Spbb{l_2,d}\Spbb{c,l_1}+\Spbb{d,l_1}\Spbb{c,l_2}\right)\frac{\Spaa{l_2,x_r}\Spaa{c,x_r}}{\Spaa{l_1,x_r}\Spaa{d,x_r}}T[\mathscr{C}(l_1)],
\eea
which precisely vanishes due to the Schouten identity (\ref{Eq:SpinorProp2}). Thus we conclude that the graphs \figref{Fig:Cancellation4} (a), (b) and (c) cancel with each other.

\begin{figure}
\centering
\includegraphics[width=0.97\textwidth]{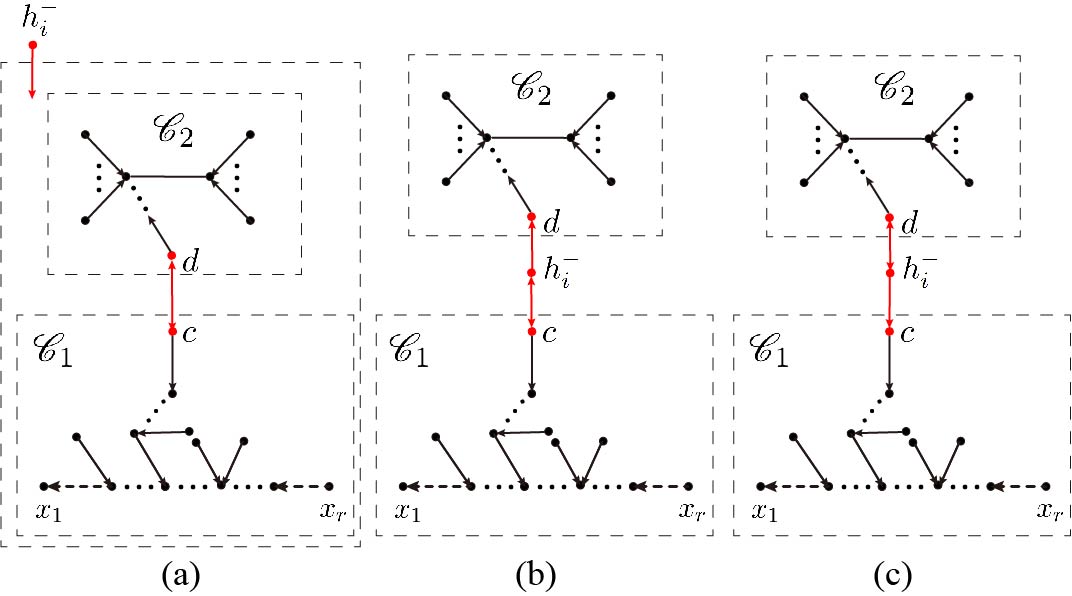}
\caption{Typical graphs involving a type-1 line for a single-trace amplitude with the $(h_i^-,h_j^-)$ configuration}\label{Fig:Cancellation5}
\end{figure}

\subsection{The vanishing amplitudes}\label{sec:VanishingAmplitude1}
Having the above properties, we are able to prove the vanishing of all the remaining amplitudes with two negative-helicity particles.
\subsubsection*{Single-trace amplitudes with the $(h_i^-,h_j^-)$ configuration}

When evaluating a single-trace amplitude with the $(h_i^-,h_j^-)$ configuration, we have to investigate not only \emph{ (i). graphs in the MHV sector which do not involve the type-3 line} but also \emph{(ii). graphs with one type-1 line} (i.e. graphs in the NMHV sector). This comes from our choice of gauge: those graphs involving the type-1 line $ \epsilon^-_{h_j}\cdot \epsilon^+$ do not vanish. We now show cancellations occur in each class of graphs.

{\bf {(i). The cancellation of graphs in the MHV sector }} In a graph with no type-1 component, the graviton $h_i$ cannot be the end node of any type-2 line that starts at a positive-helicity graviton because $\epsilon^+\cdot k_{h_i}=0$ under our choice of gauge. Hence $h_i$ can only be \emph{(1).} a leaf or \emph{(2).} an end node of a type-2 line that starts at the other negative-helicity graviton $h_j$. According to property-1, graphs in the former case are further reduced to graphs with $h_i$ removed which have the pattern \figref{Fig:Cancellation1} (a). Graphs in the latter case have the pattern \figref{Fig:Cancellation1} (b). Therefore, all graphs cancel in pairs due to property-2.

{\bf {(ii). The cancellation of graphs with one type-1 line $ (\epsilon^-_{h_j}\cdot \epsilon^+)$ }}~~In this case, we have two components that are separated by a type-3 ($k\cdot k$) line (see \appref{app:RefinedNkMHV}): the component invloving the type-1 line $ (\epsilon^-_{h_j}\cdot \epsilon^+)$ and the component involving the trace. According to our choice of gauge, the graviton $h_i$ can be either a leaf (see \figref{Fig:Cancellation5} (a)) or an end node of the type-3 line between the two components (see \figref{Fig:Cancellation5} (b) and (c)). The node $h_i$ cannot be the end node of a type-2 line that starts at $h_j$ because $\epsilon_{h_j}$ is already occupied by the type-1 line $(\epsilon^-_{h_j}\cdot \epsilon^+)$ in this case. Graphs \figref{Fig:Cancellation5} (a), (b) and (c) agree with the pattern of graphs \figref{Fig:Cancellation3} (a), (b) and (c) (when $\mathscr{C}$ in \figref{Fig:Cancellation3} includes only the graviton $h_j$ and $\mathscr{C}_2$ denotes the component involving the type-1 line), thus they cancel out due to property-3.

\subsubsection*{Double-trace amplitudes with the $(h_i^-,g_j^-)$ configuration}\label{sec:VanishingAmplitude2}
Now we turn to double-trace amplitudes with the $(h_i^-,g_j^-)$ configuration, all graphs for this case are graphs in the MHV sector, as shown by \figref{Fig:MHVSingleDouble} (b). According to our choice of gauge, $h_i$ cannot be the end node of a type-2 line which starts at a positive-helicity graviton. Hence $h_i$ can only be a leaf or an end node of the type-3 line. The typical graphs can be obtained via replacing the $\mathscr{C}_2$ in \figref{Fig:Cancellation5} (a), (b) and (c) by the component (which is defined by \figref{Fig:Components} (b) in \appref{app:RefinedNkMHV}) involving the trace $\pmb{2}$. Since the patterns of graphs  \figref{Fig:Cancellation5} (a), (b) and (c) are preserved, they must cancel out due to property-3. Thus we conclude that an amplitude with the $(h_i^-,g_j^-)$ configuration has to vanish.

\subsubsection*{Double-trace amplitudes with the $(h^-_i,h^-_j)$ configuration}
When we evaluate a double-trace amplitude with the $(h_i^-,h_j^-)$ configuration, a graph may or may not involve a type-1 line of the form $(\epsilon^-_{h_j}\cdot\epsilon^+)$. This is analogue to the single-trace case with the $(h^-_i,h^-_j)$ configuration. We state that graphs in each case cancel out.

{\bf (i). The cancellation of graphs with no type-1 line (i.e. graphs in the MHV sector, as shown by \figref{Fig:MHVSingleDouble} (b))}~~~All graphs in this case can be classified according to positions of the negative-helicity graviton $h_j$ (and the tree structure attached to it). Specifically,
$h_j$ can be an outer node or an inner node which lives on the bridge between the two traces. The former case is described by \figref{Fig:Cancellation3} (a) while the latter case has the pattern \figref{Fig:Cancellation3} (b) or (c). Here, the $\mathscr{C}_2$ component is the component which involves the trace $\pmb{2}$ (see \figref{Fig:Components} (b)). According to property-3, all these graphs cancel with each other.

{\bf (ii). The cancellation of graphs involving a type-1 line of the form $\epsilon_{h_j}^-\cdot\epsilon^+$ (i.e. graphs in the NMHV sector)}~~According to the refined graphic rule in \appref{app:RefinedNkMHV}, each graph in this case involves three components with the patterns \figref{Fig:Components} (a), (b) and (c) which respectively contain the type-1 line $\epsilon_{h_j}^-\cdot\epsilon^+$, the trace $\pmb{2}$ and the trace $\pmb{1}$. These graphs can be classified according to whether the type-1 line is on the bridges between the two traces or not. When the common kinematic factors  $\epsilon\cdot k$ and $\epsilon_{h_j}^-\cdot\epsilon^+$ in the component containing the type-1 line are extracted out, graphs in this case precisely match with the patterns \figref{Fig:Cancellation4} (a), (b) and (c) (with the correct arrows), while $\mathscr{C}_1$ and $\mathscr{C}_2$ respectively denote the components containing traces $\pmb{1}$ and $\pmb{2}$. Then we conclude that these graphs cancel with each other due to property-4.

\subsubsection*{Amplitudes with more than two traces}
For amplitudes with more than two traces and arbitrary two negative-helicity particles, there exist at least three components which are connected together by type-3 lines. In order to avoid a tedious discussion, we just sketch the main pattern of the cancellation of amplitudes in this case. As stated in \cite{Du:2019vzf}, an amplitude with more than two traces can be obtained by {\bf(i).} constructing so-called the upper and lower blocks, the latter involves the trace $\pmb{1}$, {\bf(ii).} attaching a substructure $\mathscr{C}$, which is divided by the \emph{kernel of either a type-IA or a type-IB component} (see \figref{Fig:Components} (a), (b)), to either the upper or the lower block constructed in the previous step, {\bf(iii).} connecting the two disconnected subgraphs obtained in the previous step via a type-3 line.  In a graph constructed by (i)-(iii), the kernel of the substructure $\mathscr{C}$  may be or may not be on the bridge between the upper and lower blocks.  Correspondingly,  when the kinematic factor of $\mathscr{C}$ is extracted out, such a graph has the pattern  \figref{Fig:Cancellation4} (b), (c) or (a), accompanied by the correct sign. Therefore, all graphs with the same $\mathscr{C}$ and the same configuration of upper and lower blocks must cancel out due to property-4.

\section{Conclusions}\label{sec:Conclusions}
In this paper, we evaluated all tree level  EYM amplitudes with two negative-helicity particles and an arbitrary number of traces in four dimensions, by the use of refined graphic rule.
According to the number of $\epsilon\cdot\epsilon$ lines in the graphs, all graphs were classified into $\text{N}^{k}\text{MHV}$ sectors. We pointed that the nonvanishing amplitudes with $k+2$ negative-helicity particles could at most  get contributions from graphs in the $\text{N}^{k'\,(k'\leq k)}\text{MHV}$ sectors, under a proper choice of gauge. For single-trace amplitudes with the $(g^-_i,g^-_j)$ and the $(h^-_i,g^-_j)$ configurations, we established the correspondence between the refined graphs and the spanning forests in four dimensions. A symmetric formula of double-trace amplitudes with the $(g^-_i,g^-_j)$ configuration was further provided. In this formula, the two gluon traces are on an equal footing while all gravitons are disentangled from the Parke-Taylor factors corresponding to gluon traces. Other amplitudes with two negative-helicity particles were proven to vanish, due to properties of the graphs. We now provide the following two related topics that deserve further study to end this paper.
\begin{itemize}
\item [(i)] How to extend the discussions to amplitudes with more than two negative-helicity particles?  Since the refined graphic rule is independent of dimensions, the discussion of this paper can be extended to configurations with more negative-helicity particles. Nevertheless, the general formula of $\text{N}^{k\,(k>2)}\text{MHV}$ amplitudes in  YM depends on the position of negative-helicity particles \cite{Dixon:2010ik}. This may require more hidden properties of the refined graphs in four dimensions.
\item [(ii)] It is worth studying amplitudes with gravitons coupling to matters in other theories. As pointed in \cite{Plefka:2018zwm,Zhou:2020mvz}, the recursive expansion of EYM amplitudes could be generalized to other theories with gravitons coupling to matters. Hence we expect that the calculations in the current paper, by the use of refined graphic rule (which is based on the recursive expansion of EYM amplitudes), can also be extended to these theories.
\end{itemize}

\section*{Acknowledgments}
We would like to thank Linghui Hou, Jingpei Liu and Konglong Wu for helpful discussions. This work is supported by NSFC under Grant No. 11875206, Jiangsu Ministry of Science and Technology under contract BK20170410 as well as the ``Fundamental Research Funds for the Central Universities".

\appendix

\section{Sectors in the graphic expansion of EYM amplitudes}\label{app:RefinedNkMHV}
In this part, we present the refined graphic rule for the $\text{N}^{k}$MHV sector in the expansion (\ref{Eq:PureYMExpansion1}). This rule follows from a discussion  parallel with those in \cite{Du:2017gnh,Du:2019vzf}.

\subsection{The construction of a graph $\mathcal{F}\in \mathcal{F}{[k,m]}$ in the $\text{N}^{k}$MHV sector}

To construct a graph $\mathcal{F}\in \mathcal{F}{[k,m]}$ in the $\text{N}^{k}$MHV sector, we introduce the set of all gravitons and the gluon traces $\pmb{2},\dots,\pmb{m}$, ${\pmb{\mathcal{H}}}=\{h_1,h_2,\dots,h_s,\pmb{2},\dots,\pmb{m}\}$. Each element in this set is denoted by $\mathcal{H}_{i}$ $(i=1,\dots,m+s-1)$.
We further define a reference order $\mathsf{R}$ by the ordered set
\bea
\mathsf{R}=\left\{\mathcal{H}_{\rho(1)},\mathcal{H}_{\rho(2)},\dots,\mathcal{H}_{\rho(l=m+s-1)}\right\},\Label{Eq:ReferenceOrder}
\eea
where each gluon trace is considered as a single object. The position of an element in the reference order $\mathsf{R}$ is called its \emph{weight}. When all gravitons and gluons are treated as nodes, a graph $\mathcal{F}{[k,m]}$ is obtained by connecting lines (see \figref{Fig:LineStyles}) between these nodes in an appropriate way. This can be achieved by the following two steps. \emph{\bf(i).} Construct a \emph{skeleton} that does not involve any type-3 line and may contain more than one mutually disjoint maximally connected subgraphs (which are defined as \emph{components}). \emph{\bf (ii)}. Connect these components of a skeleton via type-3 lines such that the skeleton becomes a fully connected tree graph $\mathcal{F}{[k,m]}$. Now we look into details of these two steps respectively.

\begin{figure}
	\centering
	\includegraphics[width=0.8\textwidth]{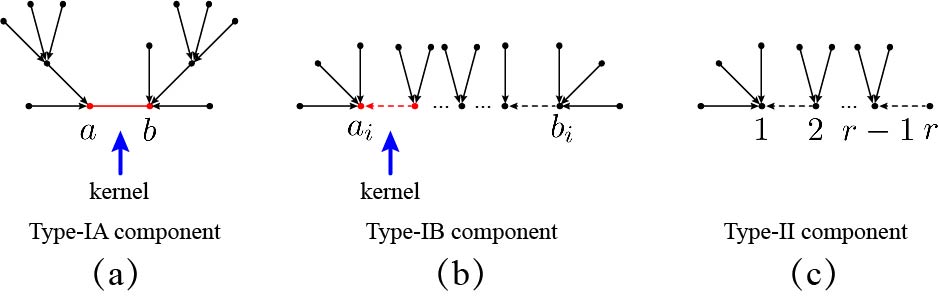}
	\caption{Three types of components defined in the refined graphic rule}\label{Fig:Components}
\end{figure}

\textbf{Step-1}~{\bf Constructing a skeleton}~~To obtain a skeleton, we first connect type-4 lines between adjacent nodes in each trace. For the trace $\pmb{1}$, the nodes are arranged in the relative order $1,2,\dots,r$ and all type-4 lines in $\pmb{1}$  point towards the direction of the node $1$.
For the trace $\pmb{i}$ ($\pmb{i}\neq \pmb{1}$), the gluons therein are arranged in a relative order
    \bea
    a_i,\beta_1,\dots,\beta_{|\pmb{i}|-2},b_i, ~~(\{\beta_1,\dots,\beta_{|\pmb{t}_i|-2}\}\equiv \pmb{\beta}\in \mathsf{KK}[\pmb{i}, a_i, b_i]). \Label{Eq:PermutationTrace}
    \eea
where $\mathsf{KK}[\pmb{i}, a_i, b_i]=\pmb{X}\shuffle \pmb{Y}^T$ when the trace $\pmb{i}$ is supposed to have the form $a_i,\pmb{X},b_i,\pmb{Y}$. All the type-4 lines in $\pmb{i}$ point towards the gluon $a_i$. We then pick out $k$ pairs of gravitons and connect a type-1 line between each pair.
A type-1 line between two gravitons or the type-4 line which is attached to the ending gluon $a_i$ of a trace is called a \emph{kernel}. All other gravitons are connected to either gluons (the gluon $r\in\pmb{1}$ is excluded) in traces or the graviton pairs with a type-1 line between them, through type-2 lines. Respectively, the arrows of these type-2 lines point towards the trace or the graviton pairs. Then a skeleton generally consists of the following two types of components:
\begin{itemize}
\item{\bf{Type-I components:}} \emph{Components which do not involve the trace $\pmb{1}$}~~This type of components can be further classified into type-IA and type-IB components. A type-IA component is defined by a component consisting of only gravitons (as shown by \figref{Fig:Components} (a)), while a type-IB component is defined by a component with a gluon trace $\pmb{i}$ ($\pmb{i}\neq\pmb{1}$) in it (as shown by \figref{Fig:Components} (b)). We define the\emph{ weight} of each type-IA or -IB component by the weight of its highest-weight node (the weight of a trace $\pmb{i}$ is always carried by its fixed gluon $a_i$ in (\ref{Eq:PermutationTrace})). Noting that each type-IA or -IB component is divided into two parts by the kernel, we further define the part containing the highest-weight node as the \emph{top side}, while the opposite part the \emph{bottom side}.

\item{\bf{Type-II component:}} \emph{Components involving the trace $\pmb{1}$}~~The typical structure of this component is shown by \figref{Fig:Components} (c). Note that the gluon $r$ cannot be connected by any type-2 line.
\end{itemize}
Since a graph $\mathcal{F}$ in the $\text{N}^{k}$MHV sector for an $m$-trace amplitude has $k$ type-1 lines, its skeleton must involve  $k$ type-IA components, $m-1$ type-IB components and one type-II component.

\textbf{Step-2}~{\bf Constructing graphs corresponding to a skeleton}~~Having a skeleton with $k$ type-IA components, $m-1$ type-IB components and one type-II component, we need to connect these components into a fully connected graph $\mathcal{F}$. This can be achieved as follows:
\begin{itemize}
\item [{\bf(i).}] For a given skeleton, we define \emph{the reference order $\mathsf{R}_{\mathscr{C}}$ of all type-I components (including type-IA and type-IB components)} by the relative order of the highest-weight nodes therein.  We further define the \emph{root set $\mathcal{R}\setminus\{r\}$} by the nodes (except the gluon $r\in \pmb{1}$) in the type-II component.

\item [{\bf(ii).}]  Supposing the reference order of components is $\mathsf{R}_{\mathscr{C}}=\left\{\mathscr{C}_1,\mathscr{C}_2,\dots,\mathscr{C}_N\right\}$, we pick out the highest-weight component $\mathscr{C}_N$ and some other components $\mathscr{C}_{i_1}$, $\mathscr{C}_{i_2}$, ..., $\mathscr{C}_{i_j}$  (which are not necessarily in the relative order in $\mathsf{R}_{\mathscr{C}}$), then construct a chain that starts from $\mathscr{C}_N$, passes through
    $\mathscr{C}_{i_j}$, $\mathscr{C}_{i_{j-1}}$, ..., $\mathscr{C}_{i_1}$ and ends at a node belonging to $\mathcal{R}\setminus\{r\}$ as follows:
    \bea
    \mathbb{CH}=\left[(\mathscr{C}_N) _t,(\mathscr{C}_N)_b\leftrightarrow(\mathscr{C}_{i_j})_{t\,(\text{or}\,b)},(\mathscr{C}_{i_j})_{b\,(\text{or}\,t)}\leftrightarrow\cdots\leftrightarrow(\mathscr{C}_{i_1})_{t\,(\text{or}\,b)},(\mathscr{C}_{i_1})_{b\,(\text{or}\,t)}\leftrightarrow\mathcal{R}\setminus\{r\}\right],
    \eea
    where we have used the subscripts $t$, $b$ to denote the top and bottom sides respectively. The notation $\leftrightarrow$ stands for the type-3 line between two nodes living in the corresponding regions. Redefine the reference order by removing those components which have been used
    \bea
    \mathsf{R}_{\mathscr{C}}\to \mathsf{R}_{\mathscr{C}}\setminus\{\mathscr{C}_N,\mathscr{C}_{i_1},\mathscr{C}_{i_2}, ..., \mathscr{C}_{i_j}\}
    \eea
    and redefine the root set by
    \bea
    \mathcal{R}\setminus\{r\}\to (\mathcal{R}\setminus\{r\})\cup \mathscr{C}_{i_1}\cup\dots\cup\mathscr{C}_{i_j}\cup\mathscr{C}_{N}.
    \eea

\item [{\bf(iii).}] Repeating the above step iteratively until the ordered set $\mathsf{R}_{\mathscr{C}}$ becomes empty, we get a fully connected tree graph $\mathcal{F}\in \mathcal{F}[k,m]$ which is rooted at the gluon $1\in\pmb{1}$.
\end{itemize}
After summing over all possible graphs $\mathcal{F}$ including (a). all choices of $a_i\neq b_i$ for fixed $b_i$ for all traces $\pmb{i}\neq \pmb{1}$, (b).  all possible permutations $\pmb{\beta}_i\in \mathsf{KK}[\pmb{i}, a_i, b_i]$ for a given trace $\pmb{i}$ and (c). all skeletons constructed for given $a_i$ and $\pmb{\beta}_i$ and (d). all graphs corresponding to a given skeleton, we get the $\text{N}^{k}$MHV sector in \eqref{Eq:PureYMExpansion1}.

\subsection{Coefficient, sign and permutations associated to $\mathcal{F}$}
The coefficient $\mathcal{C}^{\mathcal{F} }$ corresponding to a graph $\mathcal{F}$ is straightforwardly given by the product of factors corresponding to all lines in it (see \figref{Fig:LineStyles}). The sign $(-)^{\mathcal{F}}$ in \eqref{Eq:PureYMExpansion1} is defined as follows: (a). Each arrow pointing away from the gluon $1\in\pmb{1}$ contributes a minus. (b). For any type-IB component, if the kernel points away from the root (i.e. the gluon $1\in \pmb{1}$), an extra minus must be dressed. (c). Each gluon trace $\pmb{i}$ ($\pmb{i}\neq \pmb{1}$) is accompanied by $(-1)^{|\pmb{i},a_i,b_i|}$, where $|\pmb{i},a_i,b_i|$ denotes the number of elements in $\pmb{Y}$ if the trace has the form $\pmb{i}=\{a_i,\pmb{X},b_i,\pmb{Y}\}$.

Permutations $\pmb{\sigma}\in{\mathcal{F}|_{1}\setminus\{1,r\}}$ in \eqref{Eq:PureYMExpansion1} are determined as follows: (a) For two adjacent nodes $a$ and $b$, if $a$ is nearer to $1$ than $b$, we have $a\prec b$. (b). If two branches are connected to a same node, we shuffle the relative orders of the nodes belonging to these two branches.

\subsection{Graphs in the MHV sectors of single- and double-trace amplitudes}
According to the general rule, a graph $\mathcal{F}\in{\mathcal{F}{[0,1]}}$ in the MHV sector of a single-trace EYM amplitude has the pattern \figref{Fig:MHVSingleDouble} (a) which involves only one component: the type-II component. These graphs    at single-trace level are independent of the choice of reference order because only graphs with more than one components rely on the choice of reference order.

When we choose the reference order as
\bea
\mathsf{R}=\left\{h_1,h_2,\dots,\pmb{2}\right\},\Label{Eq:ReferenceOrder}
\eea
i.e., the trace $\pmb{2}$ is the highest-weight element, a typical graph $\mathcal{F}\in{\mathcal{F}{[0,2]}}$ in the MHV sector of a double-trace amplitude is then presented by \figref{Fig:MHVSingleDouble} (b). In \figref{Fig:MHVSingleDouble} (b), there is a path between the ending node $a$ of the trace $\pmb{2}$  and a gluon $l\in\pmb{1}\setminus\{x_r\}$. On this path, there exists one type-3 line and possible type-2 lines. The arrows of the type-2 lines above (or below) the type-3 line point towards the node $a\in \pmb{2}$ (or $l\in\pmb{1}\setminus\{x_r\}$). Other gravitons that do not live on the path from $a$ to $l$, can be connected to (i) gravitons on this path, or (ii) gluons of $\pmb{1}\setminus\{x_r\}$ and $\pmb{2}$, through type-2 lines whose arrows point towards the direction of the root $x_1$.

\section{Spinor helicity formalism and helpful identities}\label{app:SpinorIdentity}
We now review useful properties of spinor-helicity formalism, the Parke-Taylor formula and the Hodges formula. Helpful identities are displayed and proved in this section.

\subsection{Spinor helicity formalism}

In spinor-helicity formalism, momentum $k^{\mu}_i$ of an on-shell massless particle $i$ are mapped to two copies of two-component Weyl spinors $\lambda_i^{a}\W{\lambda}_i^{\dot{a}}$. Polarizations for  negative- and  positive-helicity gluons are respectively expressed as
\bea
\epsilon_i^{-}(\xi)\to\frac{\lambda_i^{a}\W{\lambda}_{\xi}^{\dot{a}}}{\Spbb{\xi,i}},~~~~~~~~~~\epsilon_i^{+}(\xi)\to\frac{\lambda^a_\xi\W{\lambda}_i^{\dot{a}}}{\Spaa{\xi,i}}, \Label{Eq:SpinorPolarizations}
\eea
where a normalization factor $\sqrt{2}$ (see e.g. \cite{Feng:2011np})  has been absorbed for convenience.
In the above expression, $\xi^{\mu}$ is the reference momentum which reflects the choice of gauge and the spinor products are defined by
\bea
\Spaa{i,j}\equiv \epsilon_{a{b}}\lambda_i^{a}{\lambda}_{j}^{{b}},~~~~~~~~~~\Spbb{i,j}\equiv \epsilon_{\dot{a}\dot{b}}\W\lambda_i^{\dot{a}}\W{\lambda}_{j}^{\dot{b}}, \Label{Eq:SpinorProducts}
\eea
where $\epsilon_{{a}{b}}$ and $\epsilon_{\dot{a}\dot{b}}$ are totally antisymmetric tensors. With this expression, the Lorentz contraction of two vectors are given by:
\bea
k_a\cdot k_b=\frac{1}{2}\Spaa{a,b}\Spbb{b,a},~~~k_a\cdot \epsilon^{-}_b(q)=\frac{\Spaa{a,b}\Spbb{q,a}}{\Spbb{q,b}},~~~k_a\cdot \epsilon^{+}_b(q)=\frac{\Spbb{a,b}\Spaa{q,a}}{\Spaa{q,b}}. \Label{Eq:LorentzContractions}
\eea
Helpful properties in spinor-helicity formalsim are displayed as follows.
\begin{itemize}
\item Antisymmety of the spinor products:
\bea
\Spaa{a,b}=-\Spaa{b,a},~~~\Spbb{c,d}=-\Spbb{d,c}~~~~~\Label{Eq:SpinorProp1}
\eea
When $a=b$ or $c=d$, we have $\Spaa{a,a}=0$ and $\Spbb{c,c}=0$.
\item Schouten identity:
\bea
\Spaa{a,b}\Spaa{c,d}=\Spaa{a,c}\Spaa{b,d}+\Spaa{b,c}\Spaa{d,a},~~~ \Spbb{a,b}\Spbb{c,d}=\Spbb{a,c}\Spbb{b,d}+\Spbb{b,c}\Spbb{d,a}\Label{Eq:SpinorProp2}
\eea
\item As a result of Schouten identity, we have the eikonal identity
\bea
\Sl_{i=j}^{k-1}\frac{\Spaa{i,i+1}}{\Spaa{i,q}\Spaa{q,i+1}}=\frac{\Spaa{j,k}}{\Spaa{j,q}\Spaa{q,k}}.~~~~~~~\Label{Eq:SpinorProp3}
\eea
\item Momentum conservation for an $n$-point amplitude:
\bea
\Sl_{\substack{i\neq \,j,k\\i=1}}^n\Spbb{j,i}\Spaa{i,k}=0.~~~~\Label{Eq:SpinorProp4}
\eea

\end{itemize}

\subsection{Useful identities for Parke-Taylor factors}

As a consequence of the eikonal identity, Parke-Taylor factors satisfy the following property
\bea
\Sl_{\pmb{\sigma}}\frac{1}{(1,\dots,j,\pmb{\sigma}\in\{j+1,\dots,r-1\}\shuffle\{a\},r)}&\equiv&\Sl_{\pmb{\sigma}}\frac{1}{(1,\{2,\dots,r-1\}\shuffle\{a\}|_{j\prec a},r)}\nn
&=&\frac{1}{(1,2,\dots,r)} \frac{\Spaa{j,r}}{\Spaa{j,a}\Spaa{a,r}},\Label{Eq:PTRelation1}
\eea
where $a$ is extracted out from the Parke-Taylor factors. Property (\ref{Eq:PTRelation1}) is further extended to the following more general property
\bea
\Sl_{\pmb{\sigma}}\frac{1}{(1,\dots,j,\pmb{\sigma}\in\{j+1,\dots,r-1\}\shuffle\{a_1,a_2,\dots,a_l\},r)}&\equiv&\Sl_{\pmb{\sigma}}\frac{1}{(1,\{2,\dots,r-1\}\shuffle\{a_1,a_2,\dots,a_l\}|_{j\prec a_1},r)}\nn
&=&\frac{1}{(1,2,\dots,r)} \frac{\Spaa{j,r}}{\Spaa{j,a_1}\Spaa{a_1,a_2}\dots\Spaa{a_l,r}}\nn
&\equiv&\frac{1}{(1,2,\dots,r)} \frac{\Spaa{j,r}}{\Spaa{j,a_1,\dots,a_l,r}},\Label{Eq:PTRelation2}
\eea
in which $\{a_1,\dots,a_l\}$ are extracted out from the Parke-Taylor factors. Now we prove \eqref{Eq:PTRelation1} and \eqref{Eq:PTRelation2}.

{\bf\emph{Proof of \eqref{Eq:PTRelation1}:}} The LHS of \eqref{Eq:PTRelation1} can be rewritten as
\bea
\Sl_{i=j}^{r-1}\frac{1}{(1,\dots,i,a,i+1,\dots,r)}&=&\Sl_{i=j}^{r-1}\frac{1}{(1,\dots,i,i+1,\dots,r)}\frac{\Spaa{i,i+1}}{\Spaa{i,a}\Spaa{a,i+1}}\nn
&=&\frac{1}{(1,\dots,r)}\Sl_{i=j}^{r-1}\frac{\Spaa{i,i+1}}{\Spaa{i,a}\Spaa{a,i+1}}\nn
&=&\frac{1}{(1,\dots,r)}\frac{\Spaa{j,r}}{\Spaa{j,a}\Spaa{a,r}},
\eea
where  we have collected the coefficients for the common Parke-Taylor factor $1/(1,2,\dots,r)$ on the second line and applied the eikonal identity (\ref{Eq:SpinorProp3}) on the third line. Thus \eqref{Eq:PTRelation1} has been proved.

{\bf\emph{Proof of \eqref{Eq:PTRelation2}:}} When applying the identity (\ref{Eq:PTRelation1}) repeatedly, the LHS of \eqref{Eq:PTRelation2} becomes
\bea
&&\Sl_{\shuffle}\frac{1}{(1,\dots,j,(\{j+1,\dots,r-1\}\shuffle\{a_1,a_2,\dots,a_{l-1}\})\shuffle\{a_l\}|_{a_{l-1}\prec a_l},r)}\nn
&=&\Sl_{\shuffle}\frac{1}{(1,\dots,j,(\{j+1,\dots,r-1\}\shuffle\{a_1,a_2,\dots,a_{l-1}\}),r)}\frac{\Spaa{a_{l-1},r}}{\Spaa{a_{l-1},a_l}\Spaa{a_l,r}}\nn
&=&\dots\nn
&=&\Sl_{\shuffle}\frac{1}{(1,2,\dots,r)}\left[\prod\limits_{i=0}^{l-1}\frac{\Spaa{a_{i},r}}{\Spaa{a_{i},a_{i+1}}\Spaa{a_{i+1},r}}\right],
\eea
where $a_0\equiv j$. The product in the square brackets on the last line reads
\bea
\frac{\Spaa{j,r}}{\Spaa{j,a_{1}}{\cancel{\Spaa{a_{1},r}}}}\frac{\cancel{\Spaa{a_1,r}}}{\Spaa{a_1,a_{2}}\cancel{\Spaa{a_{2},r}}}\dots\frac{\cancel{\Spaa{a_{l-2},r}}}{\Spaa{a_{l-2},a_{l-1}}\cancel{\Spaa{a_{l-1},r}}}\frac{\cancel{\Spaa{a_{l-1},r}}}{\Spaa{a_{l-1},a_{l}}\Spaa{a_{l},r}}=\frac{\Spaa{j,r}}{\Spaa{j,a_{1},a_2,\dots,a_{l},r}}.
\eea
Hence the identity (\ref{Eq:PTRelation2}) has been proven.

When $j=1$, \eqref{Eq:PTRelation2} gives the KK relation for Parke-Taylor factors:
\bea
\Sl_{\pmb{\sigma}}\frac{1}{(1,\{2,\dots,r-1\}\shuffle \pmb{\beta},r)}=(-1)^{|\pmb{\beta}|}\frac{1}{(1,2,\dots,r-1,r,\pmb{\beta}^T)}. \Label{Eq:KK}
\eea
\subsection{Parke-Taylor formula and Hodges formula}

Color-ordered tree level YM amplitude $A^{(i,j)}(1,2,\dots,n)$ with two negative-helicity gluons $i$ and $j$ satisfies the famous Parke-Taylor \cite{Parke:1986gb} formula
\bea
A^{(i,j)}(1,2,\dots,n)=\frac{\Spaa{i,j}^4}{(1,2,\dots,n)}. \Label{Eq:MHVYM}
\eea
Single-trace EYM amplitudes with two negative-helicity particles $i$, $j$ ($i$, $j$ can be two gluons or one graviton plus one gluon) can be expressed by the following Hodges determinant form \cite{Du:2016wkt}:
\bea
A^{(i,j)}(1,2,\dots,r\Vert \mathsf{H})\sim\frac{\Spaa{i,j}^4}{(1,2,\dots,r)}\det \left[\phi_{\mathsf{H}^+}\right], \Label{Eq:MHVYM}
\eea
where a normalization factor has been neglected, $1,2,\dots,r$ are gluons in the trace $\pmb{1}$, $\mathsf{H}$ denotes the graviton set, $\mathsf{H}^+$ denotes the set of positive-helicity gravitons. In the above expression, the $|\mathsf{H}^+|\times|\mathsf{H}^+|$ Hodges matrix $\phi_{\mathsf{H}^+}$ is defined as
\begin{align}
	&\phi_{ab}=\frac{[ab]}{\langle ab\rangle}& &(a\neq b)\,,& &\phi_{aa}=-\sum_{\substack{l\in\pmb{1}\cup\mathsf{H}\\l\neq a}}\frac{[al]\langle l\xi\rangle\langle l\eta\rangle}{\langle al\rangle\langle a\xi\rangle\langle a\eta\rangle}& &(a=b)\,. \Label{Eq:HodgesMatrix}
\end{align}
The spinors $\xi$ and $\eta$ represent a gauge freedom while $\phi_{aa}$ do not depend on it.


 \bibliographystyle{JHEP}
\bibliography{EYMMHVDoubleTrace}

\end{document}